\documentclass[11pt,a4paper]{article}
\usepackage{jheppub}
\pdfoutput=1

\usepackage{comment}

\usepackage{color}
\usepackage{graphicx}
\usepackage{url}
\usepackage[usenames,dvipsnames]{xcolor}
\usepackage{dcolumn}
\usepackage{bm}
\usepackage{bbm}
\usepackage{amsmath,amssymb,amsfonts}
\usepackage{dsfont}
\usepackage[vcentermath]{youngtab}
\usepackage[all]{xy}
\usepackage{pstricks}
\usepackage{dsfont}%
\usepackage{mathtools}
\usepackage{placeins}
\usepackage{enumerate}
\usepackage{cases}
\usepackage{placeins}
\usepackage{xspace}
\usepackage{cancel} 
\usepackage{slashed}
\usepackage{natbib}
\usepackage{subcaption}
\usepackage[normalem]{ulem}

\newcommand{\beq}{\begin{eqnarray}}
\newcommand{\eeq}{\end{eqnarray}}

\newcommand{\bmp}{\noindent\begin{minipage}{16cm}}
\newcommand{\emp}{\end{minipage}\vskip 7mm} 

    \newcommand{\ii}{\mathrm{i}}

    \newcommand{\SU}{\mathrm{SU}} 
    \newcommand{\Sp}{\mathrm{Sp}}
    
    \newcommand{\Tr}{\mathrm{Tr}}

    \newcommand{\SUL}{\mathrm{SU}(2)_{\mathrm{L}}} 
    \newcommand{\SUR}{\mathrm{SU}(2)_{\mathrm{R}}} 
        \newcommand{\vw}{v_{\mathrm{w}}}

    \newcommand{\fL}{f_{\Lambda}}
    
    \newcommand{\cQ}{c_{Q}}
    \newcommand{\cL}{c_{\Lambda}}

    \newcommand{\bee}{\begin{equation}}
        \newcommand{\eee}{\end{equation}}

\def\eq#1{{Eq.~\eqref{#1}}}

\def\sec#1{{Sec.~\ref{#1}}}

\def\app#1{{App.~\ref{#1}}}

\newcommand{\st}{s_\theta}
\newcommand{\ct}{c_\theta}

\def\lsim{\mathrel{\rlap{\lower4pt\hbox{\hskip1pt$\sim$}}
    \raise1pt\hbox{$<$}}}                
\def\gsim{\mathrel{\rlap{\lower4pt\hbox{\hskip1pt$\sim$}}
    \raise1pt\hbox{$>$}}}                

\begin{document}

\title{Dark matter in (partially) composite Higgs models}
\author[a]{Tommi Alanne}
 \affiliation[a]{Max-Planck-Institut f\"{u}r Kernphysik, Saupfercheckweg 1, 69117 Heidelberg, Germany}
\author[b]{Diogo Buarque Franzosi}
 \affiliation[b]{Institut f\"ur Theoretische Physik, Universit\"at G\"ottingen,
Friedrich-Hund-Platz 1, 37077 G\"ottingen, Germany} 
\author[c]{Mads T. Frandsen}
\author[c]{Martin Rosenlyst}
 \affiliation[c]{CP$^{3}$-Origins, University of Southern Denmark, Campusvej 55, DK-5230 Odense M, Denmark}

\emailAdd{tommi.alanne@mpi-hd.mpg.de}
\emailAdd{dbuarqu@gwdg.de}
\emailAdd{frandsen@cp3.sdu.dk}
\emailAdd{rosenlyst@cp3.sdu.dk}

\abstract{
We construct composite and partially composite Higgs models with complex pseudo-Nambu--Goldstone (pNGB) dark
matter states 
from four-dimensional gauge-Yukawa theories with strongly interacting fermions. The fermions 
are partially gauged under the electroweak symmetry, and the dynamical electroweak 
symmetry breaking sector is minimal. 

The pNGB dark matter particle is stable 
due to a $\mathrm{U}(1)$ technibaryon-like symmetry, also present in the technicolor limit of the models. However,
the relic density is particle anti-particle symmetric and due to thermal freeze-out as opposed to the technicolor limit where it is typically due to an asymmetry. 

The pNGB Higgs is composite or partially composite depending on the origin of the Standard Model 
fermion masses,
which impacts the dark matter phenomenology.
We illustrate the 
important features with a model example invariant under an $\SU(4)\times \SU(2) \times \mathrm{U}(1)$ global symmetry. 

{\footnotesize  \it Preprint: 
CP$^3$-Origins-2018-031 DNRF90
}
}

\maketitle
\newpage

\section{Introduction}
\label{sec:intro}
Gauge and gauge-Yukawa theories with strongly interacting fermion sectors at, or above, the weak scale underly several 
model frameworks for dynamical electroweak symmetry breaking (EWSB).  This includes 
technicolor (TC)~\cite{Eichten:1979ah,Dimopoulos:1979es,Sannino:2004qp,Dietrich:2005jn,Foadi:2012bb} and bosonic technicolor 
(bTC)~\cite{'tHooft:1979bh,Simmons:1988fu,Samuel:1990dq,Kagan:1990az,Carone:2012cd,Alanne:2013dra}, partially composite Higgs 
(pCH)~\cite{Kaplan:1983fs,Galloway:2016fuo,Agugliaro:2016clv,Alanne:2017rrs,Alanne:2017ymh}, and composite-Higgs (CH) models~\cite{Kaplan:1983sm,Luty:2004ye,Ferretti:2013kya,Ferretti:2014qta,Cacciapaglia:2014uja}. 
These models can be realized with the same underlying four-dimensional gauge theories with fermions via different 
vacuum alignments---and by adding scalars in the bTC and pCH cases.  

In the TC limit the Higgs is an excitation of the vacuum \cite{'tHooft:1979bh,Foadi:2012bb,Belyaev:2013ida}, and analogously to QCD there can be a global $\mathrm{U}(1)$ technibaryon symmetry which allows 
for the lightest technibaryon to be an asymmetric dark matter (DM) 
candidate~\cite{Nussinov:1985xr,Barr:1990ca,Gudnason:2006yj,Ryttov:2008xe,Frandsen:2009mi,Frandsen:2011kt}. 
It must typically be particle anti-particle asymmetric because of large annihilation cross sections from the strong interactions at the weak 
scale, although symmetric candidates from thermal freeze-out~\cite{Belyaev:2010kp} or the bosonic seesaw portal production mechanism in bTC~\cite{Ishida:2016fbp} can occur. 
This $\mathrm{U}(1)$ symmetry is often anomalous under weak interactions such that Stadard Model (SM) sphalerons 
may equilibrate baryon and technibaryon numbers~\cite{Barr:1990ca} 
and address the origin of the ratio of visible to dark relic densities. 

In the CH vacuum, the Higgs is realized as a pseudo-Nambu--Goldstone boson (pNGB) with properties 
which may be tuned close to the SM Higgs by external interactions. However, the $\mathrm{U}(1)$ 
technibaryon symmetry present in the TC limit is typically lost. 
Instead, DM candidates in CH models have typically been constructed and stabilized using $Z_2$ symmetries, and the models are typically studied using purely an effective Lagrangian \cite{Frigerio:2012uc,Carmona:2015haa,Fonseca:2015gva,Ballesteros:2017xeg,Ma:2017vzm}. CH models with a preserved U(1) symmetry have also been considered; see e.g. Refs~\cite{Ma:2015gra,Cai:2018tet}, and at the level of an effective description Ref.~\cite{Balkin:2017aep} and with strongly interacting scalars~\cite{Sannino:2016sfx}.

Here we are interested in models with explicit four-dimensional descriptions in terms of elementary degrees of 
freedom that retain particles charged under a global $\mathrm{U}(1)$ symmetry in the CH vacuum. 
In particular, we are interested in models where the DM candidate charged under this stabilizing U(1) symmetry 
is a pNGB related to the dynamical symmetry breaking.
This is simply realized with 
models where only part of the fermions are gauged under the electroweak (EW) interactions for which TC limits have been 
studied in e.g. Refs~\cite{Dietrich:2006cm,Luty:2008vs,Ryttov:2008xe,Frandsen:2011kt}. We, therefore,
consider models with  
$N_1+N_2$ Weyl fermions   transforming in representations $\mathcal{R}_{1}$ and $\mathcal{R}_{2}$, 
respectively,
of a new strongly interacting  gauge group $G_{\mathrm{TC}}$. Only the $\mathcal{R}_1$ fermions are 
gauged under the SM EW gauge group and responsible for EWSB. 
For fermion partial compositeness we comment on adding a third sector of $N_3$ QCD charged fermions in order to accomodate top partners. 

The $\mathcal{R}_2$ 
sector was motivated both by achieving 
near-conformal dynamics~\cite{Dietrich:2006cm,Luty:2008vs,Ryttov:2008xe} and by allowing for 
light asymmetric DM~\cite{Frandsen:2011kt} in the TC limit.
When aligning the $\mathcal{R}_1$ sector into the CH regime, the $\mathcal{R}_2$ sector can retain 
a $\mathrm{U}(1)$-technibaryon-like 
symmetry, and thereby a composite DM candidate
which, however, must now have a particle anti-particle symmetric relic density. 

The simple model we study has $G_{\mathrm{TC}}=\SU(2)$,  $N_1=4$ with $\mathcal{R}_1$ being the 
fundamental representation, and $N_2=2$ with $\mathcal{R}_2$ being the adjoint representation. 
The TC limit of this model was studied in Ref.~\cite{Ryttov:2008xe}, 
while Ref.~\cite{Luty:2008vs}  considered 
$\mathcal{R}_2=\mathcal{R}_1$ with both inert and electroweak gauged fermions in the fundamental representation. 
The CH and pCH regimes of the  $\mathcal{R}_1$ sector 
only were studied in 
Refs~\cite{Galloway:2013dma,Cacciapaglia:2014uja} 
and~\cite{Galloway:2016fuo,Agugliaro:2016clv,Alanne:2017rrs,Alanne:2017ymh}, respectively.
Since the model is formulated explicitly 
in terms of elementary constituents, the composite contributions to the spectrum may be predicted using 
lattice simulations. The $\mathcal{R}_1$ sector has recently been studied in e.g. Refs~\cite{Arthur:2016dir,Arthur:2016ozw,Pica:2016zst} and recent lattice studies have begun investigating models with multiple fermion representations \cite{DeGrand:2016mxr,Ayyar:2018zuk} and composite DM candidates~\cite{Lewis:2011zb,Hietanen:2012sz,Hietanen:2013fya,Appelquist:2014jch}.

The paper is organised as follows: In Sec.~\ref{sec:elemComp} we introduce the model framework, the model example 
with $\SU(4)\times \SU(2)\times \mathrm{U}(1)$ symmetry and the effective description. 
In Sec.~\ref{Sec:DM} we discuss the phenomenology of the  DM candidate, the pNGB of the $\mathcal{R}_2$ sector,  
and the interplay with different possible origins of the SM-fermion masses. Finally, in Sec.~\ref{sec:conclusions}
we give our conlusions.

\section{The Model and the effective description}
\label{sec:elemComp}
The model framework we propose consist of a new strongly interacting gauge group $G_{\mathrm{TC}}$ with $N_1$ Weyl fermions 
in the representation $\mathcal{R}_1$ and $N_2$ Weyl  fermions in the representation $\mathcal{R}_2$. The $\mathcal{R}_1$ 
fermions are gauged under the EW interactions, while the $R_2$ fermions are inert. 
We also add interactions to provide SM-fermion masses and to align the vacuum into the CH regime.
We will briefly discuss elementary scalars~\cite{'tHooft:1979bh} and 
four-fermion operators of both the extended-TC (ETC) ~\cite{Eichten:1979ah,Dimopoulos:1979es} and 
fermion-partial-compositeness (PC) type~\cite{Kaplan:1991dc} as examples of such interactions.
In the latter case, we add a third sector with QCD charged fermions to accomodate a top partner, to be discussed in more detail in \sec{sec:massMech}(iii).

In the minimal model example which we will study, the gauge group is chosen to be
$G_{\mathrm{TC}}=\SU(2)_{\mathrm{TC}}$ 
with $\mathcal{R}_1$ the fundamental representation and $N_1=4$. We further take $\mathcal{R}_2$ to be the 
adjoint representation and $N_2=2$ as in Ref.~\cite{Ryttov:2008xe}.  
The fermion content in terms of left-handed Weyl fields, with
$ \widetilde{\psi}_{\mathrm{L}}\equiv \epsilon \psi_R^* $,
along with their EW quantum numbers is presented in Table~\ref{tab:su4}.
\begin{table}[t]
    \caption{The new fermion content and their representations under the gauge groups, as in~\cite{Ryttov:2008xe}. 
    }
    \label{tab:su4}
    \begin{center}
	\begin{tabular}{cccc}
	    \hline
	    & $\vphantom{\frac{\frac12}{\frac12}}\quad\SU(2)_{\mathrm{TC}}\quad$ & $\SU(2)_{\mathrm{W}}\quad$ 
	    & $\mathrm{U}(1)_Y\quad$ \\
	    \hline
	    $\vphantom{\frac{\frac12}{\frac12}} (U_L,D_L)$	&   ${\tiny \yng(1)}$	&   ${\tiny \yng(1)}$	&   0\\    
	    $\vphantom{\frac{1}{\frac12}} \widetilde{U}_{\mathrm{L}}$	&   ${\tiny \yng(1)}$	&   1	&   $-1/2$\\    
	    $\vphantom{\frac{1}{\frac12}} \widetilde{D}_{\mathrm{L}}$	&   ${\tiny \yng(1)}$	&   1	&   $+1/2$ 
	    \\
	   $\vphantom{\frac{1}{\frac12}} \lambda_{ \mathrm{L}}$	&   ${\rm Adj}$	&   1	&   $0$\\    
	   $\vphantom{\frac{1}{\frac12}} \widetilde{\lambda}_{ \mathrm{L}}$	&   ${\rm Adj}$	&   1	&   $0$\\  
	    \hline
	\end{tabular}
    \end{center}
\end{table}
The strongly interacting fermion sector features the global symmetry 
$\SU(4)\times \SU(2) \times \mathrm{U}(1)$ 
at the quantum level. 
The spinors $Q=(U_{\mathrm{L}},\, D_{\mathrm{L}},\, \widetilde{U}_{\mathrm{L}},\, \widetilde{D}_{\mathrm{L}})$ 
and $\Lambda=(\lambda_{ \mathrm{L}}, \,   \widetilde{\lambda}_{ \mathrm{L}})$
transform in the fundamental representations of the $\SU(4)$ and $\SU(2)$ subgroups of the global symmetry,
respectively, where we drop a L subscript on $Q$ and $\Lambda$ for simplicity.  The anomaly free \footnote{The U(1) charges  of $Q$ and $\Lambda$ make the corresponding current anomaly free since $(-1)4\Tr[\tau^a\tau^b]+(1/2)2\Tr[\epsilon^{acd}\epsilon^{bcd}]=0$.} $\mathrm{U}(1)$ acts on both fermion sectors as 
\begin{align}
\label{eq:u1charges}
 Q \to e^{-i \alpha} Q , \quad  \Lambda \to e^{i \frac{\alpha}{2}} \Lambda\,. 
\end{align} 
In the case of PC, where we need a larger gauge group, this charge assignment is different, and we will discuss this issue in Section~\ref{sec:massMech}. 

The underlying fermionic Lagrangian we consider is 
  \begin{equation}
	\label{eq:UVLag1}
	\begin{split}
	    \mathcal{L}_{\mathrm{UV}}=&\bar{Q}\ii\slashed{D}Q+\bar{\Lambda}\ii\slashed{D}\Lambda+ \delta \mathcal{L}_m
	 + \delta \mathcal{L}\,, 
	\end{split}
    \end{equation}
where $ \delta \mathcal{L}$  are additional interactions including those responsible for vacuum alignment 
and SM-fermion masses, and we have collected the explicit mass terms for $Q,\Lambda$ in $ \delta \mathcal{L}_m$ with 
       \begin{align}
	\label{eq:MQMLambda}
 \delta \mathcal{L}_m=   \frac{1}{2}Q^T\,M_Q\,Q  +\frac{1}{2}\Lambda^T\,M_\Lambda\,\Lambda + {\rm h.c.} \, , \	M_Q=\left(\begin{array}{cc}m_1\epsilon & 0\\ 0 & -m_2\epsilon\end{array}\right)  , 
	    \ M_\Lambda=\left(\begin{array}{cc}0 & m\\ m &0 \end{array}\right).
    \end{align}
The mass terms preserve the subgroups $\Sp(4)\in \SU(4)$ and $U(1)_{\Lambda} \times Z_2 \subset \SU(2)\times \mathrm{U}(1) $.
The EW gauging of the kinetic term of $Q$ preserves the subgroup 
$\SU(2)_{\rm W} \times \mathrm{U}(1)_{Y} \times \mathrm{U}(1)_{\rm TB} \subset \SU(4)$ and the full $\SU(2) \times \mathrm{U}(1)$ part of the global symmetry. 

\subsection{The condensates and electroweak embedding}

The dynamical condensates of the theory are 
\begin{eqnarray}
\langle Q^I_{\alpha, c} Q^{J}_{\beta, c'} \epsilon^{\alpha\beta} \epsilon^{cc'} \rangle \sim  f^3 E_{Q}^{IJ} , \quad 
    \langle \lambda^{A}_{\alpha, k}\lambda^{B}_{\beta, k'} \epsilon^{\alpha\beta} \delta^{kk'}  \rangle 
    \sim  f_\Lambda^3 E_{\Lambda}^{AB} 
\end{eqnarray}
with $I,J$ and $A,B$ flavor indices in the two sectors while $\alpha, \beta$ are spin and $c,k$ are gauge indices. We expect the Goldstone boson decay constants $f,f_\Lambda$ to be of the same order~\cite{Ryttov:2008xe,Frandsen:2011kt} and will for simplicity take them to be identical later. 
The condensates break the global symmetries $\SU(4)$ to $\Sp(4)$ and $\SU(2)\times \mathrm{U}(1)$ to $\mathrm{U}(1)_{\Lambda} \times Z_2$. 
The orientation of the condensate $ E_{Q}$ relative to the EW embedding determines whether we are 
in the TC limit or in the CH regime, and the orientation is in turn determined by the interactions in $\delta \mathcal{L}, \delta \mathcal{L}_m$.  
In the TC limit the $\Sp(4)$ is aligned such that it only contains the $\mathrm{U}(1)_{\rm EM}$ subgroup 
of the electroweak symmetry group.  

To describe the general vacuum alignment in the effective Lagrangian we identify an $\SUL\times\SUR$ subgroup in 
$\SU(4)$ by the left and right generators
\begin{equation}
    \label{eq:gensLR}
    T^i_{\mathrm{L}}=\frac{1}{2}\left(\begin{array}{cc}\sigma_i & 0 \\ 0 & 0\end{array}\right),\
    T^i_{\mathrm{R}}=\frac{1}{2}\left(\begin{array}{cc} 0 & 0 \\ 0 & -\sigma_i^{T}\end{array}\right),
\end{equation}
where $\sigma_i$ are the Pauli matrices.  The EW subgroup is gauged after identifying the generator of 
hypercharge with 
$T_{\mathrm{R}}^3$; see, e.g., Refs~\cite{Luty:2004ye,Cacciapaglia:2014uja,Franzosi:2016aoo} for details. 

The alignment between the EW subgroup and the stability group $\Sp(4)$ 
can then be conveniently parameterized by an angle, $\theta$, after identifying the vacua that leave the 
EW symmetry intact, $E_Q^{\pm}$, and the one breaking it completely to $\mathrm{U}(1)_{\mathrm{EM}}$ 
of electromagnetism, $E_Q^{\mathrm{B}}$, with:
\begin{equation}
    E_{Q}^{\pm} = \left( \begin{array}{cc}
	\ii \sigma_2 & 0 \\
	0 & \pm \ii \sigma_2
    \end{array} \right)\,,\quad
    E_{Q}^B  =\left( \begin{array}{cc}
	0 & 1 \\
	-1 & 0
    \end{array} \right) \, , \quad
 E_{\Lambda}=\left( \begin{array}{cc}
	0 & 1 \\
	1 & 0
    \end{array} \right)
\end{equation}
where we have also written the $\Lambda$-sector vacuum matrix, $E_{\Lambda}$.  
The true $\SU(4)$ vacuum can be written as a linear combination of the EW-preserving and EW-breaking vacua, 
${E_Q=c_\theta E_Q^-+s_\theta E_Q^{\mathrm{B}}}$. We use the short-hand notations
$s_x\equiv \sin x, c_x\equiv \cos x$, and $t_x\equiv \tan x$ throughout.
Either choice of $E_Q^{\pm}$ is equivalent~\cite{Galloway:2010bp}, 
and here we have chosen 
$E_Q^-$. The vacuum alignment of the $\Lambda$-sector, described by the matrix $E_{\Lambda}$, is kept fixed, independent of the angle, $\theta$.

In the CH vacuum with $0<\theta<\pi/2$ the unbroken global symmetry group is reduced from Sp(4) to 
$\mathrm{U}(1)_{\rm EM} \times \mathrm{U}(1)_{\Lambda}\times Z_2$ after gauging. In the limit  $\theta=\pi/2$ referred to as the TC vacuum the unbroken symmetry group 
is $\mathrm{U}(1)_{\rm EM} \times \mathrm{U}(1)_{\rm TB} \times \mathrm{U}(1)_{\Lambda}\times Z_2$.
In the EW unbroken vacuum with $\theta=0$ it is 
$\SU(2)_{\rm W} \times \mathrm{U}(1)_{Y} \times \mathrm{U}(1)_{\Lambda}\times Z_2$.

The Goldstone excitations around the vacuum are then parameterized by  
\begin{equation}
    \begin{split}
    \Sigma_Q &= \exp\left[ 2\sqrt{2}\, i  \left(\frac{\Pi_Q}{f}-\frac{1}{3}\frac{\Theta}{f_\Theta}\mathbbm{1}_4\right) \right] E_Q, \\
    \Sigma_\Lambda &= \exp\left[ 2\sqrt{2}\,i \left( \frac{\Pi_\Lambda}{\fL}+\frac{1}{6}\frac{\Theta}{f_\Theta} \mathbbm{1}_2\right)\right] 
    E_\Lambda, \quad 
    \end{split}
\label{eq:NGBmatrix}
\end{equation}
with
\begin{equation}
    \Pi_Q=\sum_{i=1}^5 \Pi_Q^ iX^i_Q , \quad  \Pi_\Lambda=\sum_{i=1}^2 \Pi_\Lambda^ iX^i_\Lambda,
\end{equation}
where $X_{Q,\Lambda}$ are the $\theta$-dependent broken generators of SU(4) and SU(2) and can be found explicitly in Refs~\cite{Ryttov:2008xe,Galloway:2010bp}.
The $\Theta$ state is the only state connecting the two sectors at the level of single trace terms in the effective Lagrangian and its U(1) charges under each sector follow from  \eq{eq:u1charges}.

For simplicity, we will henceforth use the notations $h\equiv\Pi_Q^4$ and $\eta\equiv \Pi_Q^5$ 
for the composite Higgs, and the CP-odd pNGB of the Q sector, resp., and 
$\Phi\equiv\frac{1}{\sqrt{2}}(\Pi_\Lambda^1-i \Pi_{\Lambda}^2)$,
$\bar{\Phi}\equiv\frac{1}{\sqrt{2}}(\Pi_\Lambda^1+i \Pi_{\Lambda}^2)$ for the $\Lambda$-sector pNGBs, corresponding to $\Lambda^T C \Lambda$ and 
$\bar{\Lambda} C \bar{\Lambda}^T$ states.

The states in the EW unbroken limit with  $\st=0$  (upper) and TC vacuum with $\st=1$ (lower) are given in Table~\ref{table:symmetries}. In the composite Higgs range the states correspond to those in the EW unbroken vacuum except $h\equiv \Pi_Q^4\sim  \ct (\bar{U}U+\bar{D}D) + \st \,{\rm Re} \, U^TCD$, see e.g. Ref.~\cite{Franzosi:2016aoo}.
\begin{table}[h]
\caption{Table of pNGB states in the EW unbroken limit with $\st=0$ (upper) and the TC limit $\st=1$  (lower)
}
\begin{eqnarray}
\begin{array}{cccc}
\hline
 \vphantom{\sum_{\frac12}^{\frac12}}\qquad& \qquad \mathrm{U}(1)_{\rm TB}\qquad\quad & \qquad \mathrm{U}(1)_{\Lambda}\qquad\quad& \qquad  Z_2 \qquad\ \\  \hline
\vphantom{\sum_i^{\frac12}}h\equiv \Pi_Q^4\sim \bar{U}U+\bar{D}D  & - &  0 &  0 \\
\vphantom{\sum_i^{\frac12}}\eta\equiv \Pi_Q^5 \sim \text{Im}\,U^TCD & -& 0 &  0 \\
\vphantom{\sum_i^{\frac12}}\Phi\equiv (\Pi_\Lambda^1-i\Pi_\Lambda^2)/\sqrt{2}\sim \Lambda^T C \Lambda  & - &  1 &  0 \\
\vphantom{\sum_i^{\frac12}}\bar\Phi\equiv (\Pi_\Lambda^1+i\Pi_\Lambda^2)/\sqrt{2}\sim \bar{\Lambda} C \bar{\Lambda}^T   & - &  -1 &  0 \\
\vphantom{\sum_{\frac{\frac12}{2}}^{\frac12}}\Theta \sim i(\bar{U}\gamma^5 U+\bar{D}\gamma^5 D - (1/2)\bar{\Lambda}\gamma^5\Lambda )  & - &  0 &  0 \\
 \hline
\vphantom{\sum_{\frac12}^{\frac{\frac12}{2}}}\Pi_{UD} \equiv (\Pi^4_Q-i\Pi^5_Q)/\sqrt{2} \sim U^TCD & \frac{1}{\sqrt{2}} &  0 &  0 \\
\vphantom{\sum_i^{\frac12}}\Pi_{\overline{U}\overline{D}}\equiv (\Pi_Q^4+i\Pi_Q^5)/\sqrt{2} \sim \bar{U}C\bar{D}^T  & -\frac{1}{\sqrt{2}} &  0 &  0 \\
\vphantom{\sum_i^{\frac12}}\Phi  & 0 &  1 &  0 \\
\vphantom{\sum_i^{\frac12}}\bar\Phi  & 0 & -1 &  0 \\
\vphantom{\sum_{\frac12}^{\frac12}}\Theta & 0 &  0 &  0\\ 
\hline
\end{array} \nonumber
\end{eqnarray}
\label{table:symmetries}
\end{table}

Below the condensation scale, the Lagrangian of Eq.~\eqref{eq:UVLag1}, gauged under the EW interactions, yields
\begin{equation}
    \label{eq:effLag}
    \mathcal{L}_\mathrm{eff}=\mathcal{L}_{\mathrm{kin}}-V_{\mathrm{eff}},
\end{equation}
where the kinetic terms are
\begin{equation}
    \label{eq:kinLag}
    \mathcal{L}_{\mathrm{kin}}=\frac{f^2}{8}\Tr [D_{\mu}\Sigma_Q^{\dagger}D^{\mu}\Sigma_Q]+\frac{\fL^2}{8}\Tr [\partial_{\mu}\Sigma_\Lambda^{\dagger}\partial^{\mu}\Sigma_\Lambda], 
\end{equation}
with 
\begin{equation}
    \label{eq:covD}
    D_{\mu}\Sigma_Q=\partial_{\mu}\Sigma_Q-\ii\left(G_{\mu}\Sigma_Q+\Sigma_QG_{\mu}^{T}\right),
\end{equation}
and the EW gauge fields are encoded in the covariant derivative
\begin{equation}
    \label{eq:Gmu}
    G_{\mu}=g W_{\mu}^iT_{\mathrm{L}}^i+g^{\prime}B_{\mu}T_{\mathrm{R}}^3.
\end{equation}
    The kinetic term of $\Theta$ is canonically normalized only if $f_\Theta=f=\fL$ which we will assume for simplicity. In the general case, the kinetic terms must be renormalized but based on Casimir scaling we expect them to be of the same size~\cite{Ryttov:2008xe,Frandsen:2011kt}.

    The EW gauge interactions contribute to the effective potential at the one-loop level,
    but the contribution is higher order as compared to the vector-like mass terms in Eq.~\eqref{eq:MQMLambda}, and numerically subleading due to the smallness of the EW gauge
    couplings
    as compared to the top-loop contributions arising from the four-fermion interactions.  We discuss the latter below. 

The effective potential, at the lowest order, is given by
\begin{equation}
    \label{eq:V0}
   V_{\mathrm{eff}, m}^0=2\pi\cQ f^3\,\Tr\left[M_Q\Sigma_Q^{\dagger}+\Sigma_QM_Q^{\dagger}\right]
    +2\pi\cL \fL^3\,\Tr [M_{\Lambda}\Sigma_{\Lambda}^{\dagger}+\Sigma_{\Lambda}M_{\Lambda}^{\dagger}],
\end{equation}
where $\cQ$, $\cL$ are non-perturbative $\mathcal{O}(1)$ constants, and we use the numerical value 
${\cQ\approx 1.5}$ 
suggested by the lattice simulations~\cite{Arthur:2016dir}. 
The mass terms involving $M_Q$ (as well as the subleading EW gauge interactions) prefer the vacuum where the EW is unbroken 
as we discuss below. The correct vacuum alignment must therefore be ensured by the SM-fermion 
mass generation mechanism.

\subsection{Interactions between the $\mathcal{R}_1$ and $\mathcal{R}_2$ sectors}
From the single trace terms in Eq.~\eqref{eq:V0}, the only interactions between the $\mathcal{R}_1$ and $\mathcal{R}_2$ sectors are those involving  
the $\Theta$ state.
At the next leading order all interactions between the  $\mathcal{R}_1$ and $\mathcal{R}_2$ sectors arise 
from double trace terms
\begin{equation}
    \label{eq:intLag}
    \begin{split}
    \mathcal{L}_{Q,\Lambda}=&\frac{c_{1}}{4\pi} \Tr[D_{\mu}\Sigma_Q^{\dagger}D^{\mu}\Sigma_Q]\Tr[\partial_{\mu}\Sigma_\Lambda^{\dagger}\partial^{\mu}\Sigma_\Lambda]\\
    &-\frac{c_{2}}{4\pi} f_\Lambda  \Tr [D_{\mu}\Sigma_Q^{\dagger}D^{\mu}\Sigma_Q]\Tr[M_\Lambda \Sigma_\Lambda^{\dagger}+ \Sigma_\Lambda M_\Lambda^\dagger-2M_{\Lambda}E_{\Lambda}] \\
    &- \frac{c_{3}}{4\pi} f_\Lambda f  \Tr[M_Q \Sigma_Q^{\dagger}+ \Sigma_Q M_Q^\dagger+2M_QE_Q] \Tr[M_\Lambda \Sigma_\Lambda^{\dagger}+ \Sigma_\Lambda M_\Lambda^\dagger -2M_{\Lambda}E_{\Lambda}]\\
    &- \frac{c_{4}}{4\pi} f \Tr[M_Q \Sigma_Q^{\dagger}+ \Sigma_Q M_Q^\dagger+2M_QE_Q] \Tr[\partial_{\mu}\Sigma_\Lambda^{\dagger}\partial^{\mu}\Sigma_\Lambda] + ...,
    \end{split}
\end{equation}
where $c_i, 1=1,\dots,4$ are the  Gasser--Leutwyler type coefficients~\cite{Gasser:1984gg}, and 
$c_i\sim\mathcal{O}(1)$  by naive dimensional analysis in analogy with QCD~\cite{Manohar:1983md}.
We have for simplicity shifted the $c_2$ and $c_3$ terms such that
$\Phi, h$, and the EW gauge boson  masses and kinetic terms do not acquire additional contributions from 
these higher-order terms, but are instead determined by 
Eqs~\eqref{eq:V0},~\eqref{eq:mhcomp}, and~\eqref{eq:kinLag}, resp.

Expanding Eqs~\eqref{eq:V0} and~\eqref{eq:intLag} yields 
 \begin{equation}
 \label{eq:intLag2}
 \begin{split}
\hspace{-0.05cm}\mathcal{L}_{Q,\Lambda}\supset &
\frac{m_{\Phi}^2}{\fL^2} \Phi\bar{\Phi} \left(g_h  h  + \frac{1}{2}g_{ZZ}   Z_\mu Z^\mu +  g_{WW}   W^+_\mu W^{-\mu} 
+  \frac{1}{2}g_{\Theta\Theta} \Theta^2  + \frac{1}{2} g_{hh}  h^2+ \frac{1}{2} g_{\eta \eta} \eta^2  \right.\\
    &\left.\qquad\quad+  \frac{1}{2}g_{\partial\Theta}\, \partial_\mu \Theta  \partial^\mu \Theta
	+  \frac{1}{2}g_{\partial h}\,  \partial_\mu h\partial^\mu h+  \frac{1}{2}g_{\partial\eta}\, \partial_\mu\eta\partial^\mu\eta \right)
  \\
  & +\frac{1}{\fL^2}\partial_\mu \Phi\partial^\mu \bar{\Phi} \left(   d_h  h  +\frac{1}{2} d_{ZZ}  Z_\mu Z^\mu +  d_{WW}   W^+_\mu W^{-\mu} 
    +  \frac{1}{2}d_{hh} h^2+ \frac{1}{2}d_{\eta \eta}  \eta^2 \right.\\
    &\left.\qquad\qquad\qquad+  \frac{1}{2}d_{\partial\Theta}\, \partial_\mu \Theta  \partial^\mu \Theta
	+  \frac{1}{2}d_{\partial h}\,  \partial_\mu h\partial^\mu h+ \frac{1}{2} d_{\partial\eta}\, \partial_\mu\eta\partial^\mu\eta \right),
  \end{split}
    \end{equation}
    where
   \begin{equation}
   m_{\Phi}^2= -16\pi\cL \fL m,
\end{equation}    
analogously to the Gell-Mann--Oakes--Renner relation in QCD~\cite{GellMann:1968rz}.
The mass appears as a common
prefactor of all non-derivative terms as a consequence of the Goldstone nature of $\Phi$. The couplings involving the $\Theta$ state
$g_{\Theta\Theta}$ and $g_{\partial\Theta}$ are the only ones not arising from a double trace term. 
We give the explicit couplings in appendix~\ref{sec:intLag}.

Besides the anomaly-free $\Theta$ state, a $\Theta'$ state corresponding to the  U(1) which is quantum anomalous is also present in the spectrum of the theory and can mix with $\Theta$. Its mass is generated by instanton effects related to the U(1) anomaly~\cite{Belyaev:2016ftv}.
 We provide more details in \app{sec:u1bosons}.
We assume for simplicity that $\Theta'$ mass is large and it decouples.

The anomaly-free state $\Theta$ on the other hand receive its mass from explicit U(1) breaking terms, like the 
vector-like mass terms of Eq.~\eqref{eq:V0}, which also generate interactions between $\Theta$ and other pNGBs.
Explicitly, the relevant terms (excluding derivatives)
up to quartic order are given by
\begin{equation}
\label{eq:thetamass}
\begin{split}
&{\cal L} \supset\ \frac{8\pi}{9f_\Theta^2}\left(4\cQ f^3 m_Q \ct + \cL \fL^3 m\right)\Theta^2 
    - \frac{16\sqrt{2}\pi}{3f_\Theta} \cQ f^2 (m_1-m_2)\Theta\eta \\
& -\frac{4}{9\pi f_\Theta^2}\st m_Q(8\pi^2\cQ f^2+c_3\fL m) h \Theta^2 
    -\frac{32  m}{9\pi f_\Theta^2 f_\Lambda}(2\pi^2\cL\fL^2+c_3m_Q f\ct)\Theta^2\Phi\bar{\Phi}\\
&+\frac{2}{9\pi f f_\Theta^2}\left\{\vphantom{\frac12}(h^2+\eta^2)\left[4\sqrt{2}\pi^2 \cQ f f_\Theta(m_1-m_2) \Theta \eta 
    - \ct m_Q (8\pi^2\cQ f^2+c_3 m \fL) \Theta^2\right]\right.\\
&\left.\qquad\qquad\quad+\frac{4\sqrt{2}}{9f_\Theta} (m_1-m_2)(8\pi^2\cQ f^2+3c_3m \fL) 
    \Theta^3\eta \right\}
\end{split}
\end{equation}
where $m_Q\equiv m_1+m_2$. We also assume that $m_1=m_2$ so mass mixing is absent in the pseudoscalar sector.

The relevance of the $\Theta$ state and its interactions for our study is that the quartic term $\Phi\bar{\Phi} \Theta^2$  can erase the thermal relic density of $\Phi$ unless
$m_\Theta > m_{\Phi}$ or $m\ll f_\Lambda$. 
In the following we require $m_\Theta>m_{\Phi}$ and this imposes non-trivial constraints on the parameter space.

\subsection{SM-fermion masses and vacuum alignment}
\label{sec:massMech}

The composite sector must be extended to provide SM-fermion masses and correct vacuum alignment. Here we briefly comment 
on the three different possibilities which will impact the DM phenomenology. Further ways to distinguish these possibilities using the pseudoscalars analogous to the QCD $\eta, \eta'$ states have been discussed in Ref.~\cite{Alanne:2016rpe}.

\bigskip
{\bf (i) ETC-type four-fermion operators.} One possibility is to add four-fermion operators as in Ref.~\cite{Kaplan:1983sm}. 
Such four-fermion operators could themselves arise from the exchange of heavy scalar multiplets but also from heavy 
vectors as in ETC~\cite{Dimopoulos:1979es}. 

Explicitly for the top quark, after integrating out heavy states, we would have  
  \begin{equation}
	\label{eq:1f}
	\begin{split}
	    \mathcal{L}_{\mathrm{4f}}\sim&-\frac{Y_tY_U}{\Lambda_t^2}(\bar{q}_{\mathrm{L}}t_{\mathrm{R}})^{\dagger}_{\alpha}
		(Q^{T}P_{\alpha}Q)+\ \mathrm{h.c.},
	\end{split}
    \end{equation} 
where the spurion, $P_{\alpha}$,  projects out the EW components such that $ Q^T P_{\alpha}Q$ 
transforms as EW doublet with hypercharge $+1/2$.
Upon the condensation of the techniquarks, this yields a contribution to the top mass, i.e.
    \begin{equation}
	\label{eq:L4f}
	    \mathcal{L}_{\mathrm{4f}}\sim-4\pi \cQ f^3\frac{Y_tY_U}{\Lambda_t^2}\Tr[P_1\Sigma]\bar{t}t
		=-y_t^{\prime}f s_{\theta} \ \bar{t}t
		-y_t^{\prime}c_{\theta}\, h\,\bar{t}t+\dots,
    \end{equation} 
    where 
    \begin{equation}
	\label{eq:}
	y_t^{\prime}\equiv\frac{4\pi \cQ f^2 Y_tY_U}{\Lambda_t^2}=\frac{m_t}{\vw}.
    \end{equation}
    This yields the Higgs-top coupling
    \begin{equation}
	\label{eq:}
	y_{h\bar{t}t}=y_t^{\prime}c_{\theta}=\frac{m_t}{\vw}c_{\theta} , 
    \end{equation}
and gives a contribution to the effective potential via the top loop:
    \begin{equation}
	\label{eq:}
	V_{\mathrm{top}}=-C_t y_t^{\prime\,2}f^4\sum_{\alpha}|\Tr[P_{\alpha}\Sigma_Q]|^2
	    =-C_ty_t^{\prime\, 2}f^4s_{\theta}^2+\dots,
    \end{equation}
    where $C_t$ encodes the non-perturbative top-loop effects.
    The vacuum alignment is dominantly
    given by balancing the top contribution and contribution from the explicit techniquark mass 
    terms~\cite{Cacciapaglia:2014uja}:
    \begin{equation}
	\label{eq:alignComp}
	c_{\theta}=-\frac{4\pi \cQ m_Q}{C_t y_t^{\prime\,2}f}.
    \end{equation}
    The Higgs mass is given by 
    \begin{equation}
	\label{eq:mhcomp}
	m_h^2=-2f\left(4\pi\cQ m_Qc_\theta+C_t y_t^{\prime\,2}fc_{2\theta}\right)=2C_ty_t^{\prime\, 2}f^2s_{\theta}^2=2C_t m_t^2.
    \end{equation}
    Thus the correct Higgs mass is reproduced for $C_t\sim\frac{1}{4}$. Furthermore, the mass of $\eta$ reads
    \begin{equation}
	\label{eq:}
	m_{\eta}^2=m_h^2/s_\theta^2.
    \end{equation}

 We can write the mass of $\Theta$ from Eq.~\eqref{eq:thetamass} now as 
\begin{equation}
\label{eq:ThetaMassCH}
m_\Theta^2 = \frac{8}{9}m_\eta^2c_\theta^2 + \frac{1}{9}m_{\Phi}^2.
\end{equation}
Requiring that $m_\Theta>m_{\Phi}$ leads to the constraint
\begin{equation}
    \label{eq:constraintCH}
    m_\Phi<m_\eta c_\theta,    
\end{equation}
and further after parametrizing $m_\Phi=\epsilon f$, to a relation between $s_\theta$ and $\epsilon$:
\begin{equation}
    \label{eq:maxepsCH}
    \epsilon^2<\frac{m_h^2}{\vw^2}(1-s_\theta^2)\approx \frac{1}{4}(1-s_\theta^2).
\end{equation}

{\bf (ii) Partially composite Higgs.}
Alternatively we may add a multiplet of elementary scalars containing at least an $\SU(2)_{\mathrm{W}}$ doublet with Yukawa couplings to 
 SM fermions as in Refs~\cite{'tHooft:1979bh,Kaplan:1983fs,Galloway:2016fuo,Agugliaro:2016clv,Alanne:2017ymh}. 
 We will focus on the simplest example with $H$ an $\SU(2)_{\mathrm{W}}$ 
 doublet by adding to the UV Lagrangian above the interactions: 
 \begin{equation}
	\label{eq:UVLag2}
	\begin{split}
	    \mathcal{L}_{H}=& 
	    D_{\mu}H^{\dagger}D^{\mu}H-m_H^2H^{\dagger}H-\lambda_H(H^{\dagger}H)^2\\
	  -& y_U H_{\alpha}(Q^{T}P_{\alpha}Q)-y_D \widetilde{H}_{\alpha}(Q^{T}\widetilde{P}_{\alpha}Q)+\mathrm{h.c.},
	\end{split}
    \end{equation}
where the elementary Higgs doublet 
is given by
       \begin{align}
	\label{eq:H}
	H_\alpha&=\frac{1}{\sqrt{2}}\left(\begin{array}{c}\sigma_h-\ii \pi_h^3 \\ -(\pi_h^2+\ii \pi_h^1)\end{array}\right) , 
    \end{align}
      where   
    $\widetilde{H}=\epsilon H^*$. The antisymmetric contractions
    are kept implicit. 
    
    The above Yukawa interactions induce the following lowest-order operators to the effective potential:
    \begin{equation}
	\label{eq:}
	    V_{\mathrm{pCH}}=4\pi\cQ f^3\left(y_U H_\alpha\Tr\left[P_{\alpha}\Sigma_Q\right]
		+y_D\widetilde{H}_{\alpha}\Tr[\widetilde{P}_{\alpha}\Sigma_Q]+\ \text{h.c.}\right).
    \end{equation}
This part of the potential prefers the TC vacuum such that the final vacuum alignment in this case is given by
    \begin{equation}
	\label{eq:}
	t_{\theta}=-\frac{(y_U+y_D)v}{\sqrt{2}(m_1+m_2)}=-\frac{v\, y_Q}{\sqrt{2}\,m_Q},
    \end{equation}
    where $v\equiv \langle\sigma_h\rangle$, $y_Q\equiv y_U+y_D$ and again $m_Q = m_1 + m_2$.

The two CP-even mass eigenstates in this partially composite Higgs case are given in terms of the interaction eigenstates by~\cite{Alanne:2017ymh}
    \begin{equation}
	h_1 = c_\alpha \sigma_h - s_\alpha  h,\quad
	h_2 = s_\alpha \sigma_h + c_\alpha  h , 
  \label{Eq:Higgsmixing}
    \end{equation}    
where as above we have first identified $h\equiv\Pi^4$, and the angle, $\alpha$, is determined by
\begin{equation}
    \label{eq:alpha-delta}
    t_{2\alpha}=\frac{f v s_{2\theta}}{(1+\delta)f^2s_\theta^2-v^2},\quad\text{where}\quad
    \delta\equiv\frac{2\lambda_H v^2}{m_H^2+\lambda_H v^2}.
\end{equation}
The ordinary CH case corresponds to $s_\alpha =-1, h_1=h$.
In this case, the $\eta$ mass is given by
\begin{equation}
    \label{eq:}
    m_\eta^2=\frac{m_h^2 t_\alpha}{t_\alpha+c_\theta/t_\beta},  
\end{equation}
where the angle $\beta$ is defined by
\begin{equation}
    \label{eq:}
    t_{\beta}=\frac{v}{f s_{\theta}}.
\end{equation}

The mass of $\Theta$ can now be written as
\begin{equation}
\label{eq:ThetaMassE}
m_\Theta^2 = \frac{8}{9}m_\eta^2 + \frac{1}{9}m_\Phi^2.
\end{equation}
Requiring that $m_\Theta>m_{\Phi}=\epsilon f$ leads again to a condition between $s_\theta$ and $\epsilon$:
\begin{equation}
    \epsilon^2<\frac{m_\eta^2 }{\vw^2} s_\theta^2.
    \label{eq:maxepsE}
\end{equation}

{\bf (iii) Fermion partial compositeness.} Finally in the partial compositeness scenario we add baryon-like operators as 
in Ref.~\cite{Kaplan:1991dc}. In this case the underlying theory has to be extended with extra QCD charged fermions in order to construct the top partners and to enlarge the $G_{\mathrm{TC}}$ gauge group to ensure asymptotic freedom as studied in e.g. Ref.~\cite{Ferretti:2013kya}. 
The effect of the partial compositeness operators into vacuum alignment was extensively discussed in Ref.~\cite{Alanne:2018wtp}. The vacuum alignment phenomemnology depends on in which  
representation of the global symmetry group the top partners are embedded. For concreteness, we will 
consider the top partners in the symmetric representation of $\SU(4)$.

The top mass and linear couplings to pNGBs can be written as~\cite{Alanne:2018wtp}
\begin{equation}
    \begin{split}
     &\frac{C_{yS}}{4\pi} y_{t_L} y_{t_R}f (Q_1 t^c)^\dagger~{\rm Tr} [P_Q^1 \Sigma^\dagger P_t\Sigma^\dagger] +\mathrm{h.c}\\
    &=(Q_1 t^c)^\dagger \left( m_\mathrm{top} + \frac{m_\mathrm{top}}{v} \left( \frac{c_{2\theta}}{c_\theta}\ h 
	- i \frac{s_\theta}{c_\theta}\ \eta \right) + \dots \right) +\mathrm{h.c}\,, 
    \end{split}
\end{equation}
with $C_{yS}\sim\mathcal{O}(1)$,  $m_t = \frac{C_{yS}}{4\pi} y_{t_L} y_{t_R} c_\theta s_\theta f$,  
and the contribution to the effective potential is given by
\begin{equation}
\begin{split}
V_{\mathrm{PC}}=& \frac{C_{tS}}{(4\pi)^2} f^4 \left( y_{t_L}^4 {\rm Tr} [P_Q^\alpha \Sigma^\dagger P_Q^\beta
    \Sigma^\dagger]{\rm Tr}[\Sigma P^\dagger_{Q \alpha} \Sigma P^\dagger_{Q\beta}]\right.\\
&\qquad\qquad +y_{t_R}^4 ~{\rm Tr} [P_t \Sigma^\dagger P_t
    \Sigma^\dagger]{\rm Tr}[\Sigma P^\dagger_{t} \Sigma P^\dagger_{t}] \\
&\qquad\qquad+
    \left.   y_{t_L}^2 y_{t_R}^2 ~{\rm Tr} [P_Q^\alpha \Sigma^\dagger P_t
    \Sigma^\dagger]{\rm Tr}[\Sigma P^\dagger_{Q \alpha} \Sigma P^\dagger_t] \right)\\
=& \frac{C_{tS}}{(4\pi)^2}  f^4 ~ \left( y_{t_L}^4 s_{\theta}^4 + y_{t_R}^4 c_{\theta}^4 + y_{t_L}^2 y_{t_R}^2 
    c_{\theta}^2  s_{\theta}^2 \right) ,
    \end{split}
\end{equation}
where
\begin{equation}
\begin{split}
P_{Q}^1=\dfrac{1}{\sqrt{2}}
\begin{pmatrix}
0 & 0 & 1 &0
\\
0 & 0 &0 & 0
\\
1 & 0 & 0 &0
\\
0 & 0 & 0 &0
\end{pmatrix}\,,
\quad
P_{Q}^2=\dfrac{1}{\sqrt{2}}
\begin{pmatrix}
0 & 0 & 0 &0
\\
0 & 0 & 1 & 0
\\
0 &  1 & 0 &0
\\
0 & 0 & 0 &0
\end{pmatrix}\,, 
\quad
 P_{t}=\dfrac{1}{\sqrt{2}}
\begin{pmatrix}
0 & 0 & 0 &0
\\
0 & 0 &0 & 0
\\
0 & 0 & 0 & 1
\\
0 & 0 &  1 &0
\end{pmatrix}\,,
\end{split}
\end{equation}
and $C_{tS}\sim\mathcal{O}(1)$.

Now the alignment condition reads
\begin{equation}
    \label{eq:alignPC}
    m_Q=\frac{4\pi C_{tS}}{c_QC_{yS}^4y_{t_R}^4f^3}
    \left[ \frac{m_t^2}{c_\theta^3s_\theta^2}
    \left(\frac{C_{yS}^2}{(4\pi)^2}y_{t_R}^4 f^2 c_{\theta}^2c_{2\theta}+2m_t^2\right)
    -2\frac{C_{yS}^4}{(4\pi)^4}y_{t_R}^8f^4 c_{\theta}^3\right],
\end{equation}
and the Higgs mass is given by
\begin{align}
    \label{eq:}
    m_h^2=&\frac{(4\pi)^2C_{tS}}{C_{yS}^4 y_{t_R}^4f^2}\left[
	\frac{C_{yS}^2}{(4\pi)^2}f^2y_{t_R}^4\left(\frac{3C_{yS}^2}{(4\pi)^2}y_{t_R}^4f^2s_{2\theta}^2
    +2\left(t_\theta^2-5\right)m_t^2\right)
	+m_t^4\left(16/s_{2\theta}^2-2/c_{\theta}^4\right)
    \right]\nonumber\\
    =&\frac{32\pi^2C_{tS}}{C_{yS}^4}\left(\frac{4m_t^4}{y_{t_R}^4\vw^2}-\frac{5C_{yS}^2}{(4\pi)^2}m_t^2
	+\frac{6C_{yS}^4}{(4\pi)^4}y_{t_R}^4\vw^2\right)+\mathcal{O}(s_\theta^2)\,.
\end{align}
Obtaining the correct Higgs mass, requires a small hierarchy $C_{tS}< C_{yS}$.
Finally, the $\eta$ mass can be written as
\begin{equation}
    \label{eq:}
    m_\eta^2=\frac{4C_{tS}}{(4\pi)^2}f^2y_{t_R}^4c_\theta^2-\frac{8\pi^2C_{tS}m_t^2}{C_{yS}^4f^2y_{t_R}^4s_\theta^2c_\theta^4}
	\left(\frac{C_{yS}^2}{(4\pi)^2}f^2y_{t_R}^4(c_{4\theta}+2 c_{2\theta}+1)
	+8m_t^2\right).
\end{equation}

One possible extension of our model    
to provide the top partners
is to have four Weyl fermions, $Q$, in the fundamental ({\bf F}), another six, $\chi$,  in the 
two-index anti-symmetric, ${\bf A_2}$, as detailed in Ref.~\cite{Ferretti:2014qta}, and keep the inert sector  
with $\lambda$-fermions still in the adjoint representation ({\bf G}) of $G_{\mathrm{TC}}$.
In this case, it is also necessary to modify the gauge group in order to ensure asymptotic freedom, 
one possibility being $G_{\mathrm{TC}}=\Sp(N_{\mathrm{TC}})$.
 The enhanced global symmetry is 
then SU(4)/Sp(4)$\times$ SU(2)/SO(2) $\times$ SU(6)/SO(6) $\times$ U(1)$^2$. 
Asymptotic freedom is guaranteed by $N_{\mathrm{TC}}=2<3$, where the first coefficient $b$ of the beta function is positive\cite{Ryttov:2008xe,Barnard:2013zea}, i.e.
\begin{equation}
b=\frac{11}{3}C_2(\mathrm{\bf G})-\frac{2}{3}(4T(\mathrm{\bf F})+2T(\mathrm{\bf G})+6T(\mathrm{\bf A_2}))=-\frac{10}{3}(N_{TC}-3)>0
\end{equation}
with $T(\mathcal{R})=1,\,2N_{\mathrm{TC}}-2,\,2N_{\mathrm{TC}}+2$ the index of representation $\mathcal{R}={\bf F,\,A_2,\,G}$ respectively. 
The two anomaly-free U(1) give rise to two extra pNGBs. 
The $Q$, $\chi$ and $\lambda$ charges under these U(1) are defined by the anomaly cancellation
\begin{equation}
q_Q T(F)+q_\chi T(A_2) + q_\lambda T(G)=0\Rightarrow q_\lambda = - \frac{1}{3}q_Q-q_\chi\,.
\end{equation}
The two states will mix 
but we leave a detailed study for the  future and restrict  
to the simple case where the $\chi$-sector (as well as the anomalous $\Theta'$) decouple (for instance with an explicit $\chi$ mass) so that 
the lighest $\Theta$-state can be determined by the charge assignment $q_Q=-1$, $q_\lambda=1/3$.
All interactions of the $\Theta$-sector then have to be modified accordingly;
in particular the mass relation between $\Phi$ and $\Theta$ is modified to
\begin{equation}
\label{eq:ThetaMassPC}
m_\Theta^2 = \frac{1}{4q_Q^2+2q_\lambda}(4q_Q^2 m_\eta^2c_\theta^2 + 2 q_\lambda^2 m_{\Phi}^2)
 = \frac{1}{19}(18 m_\eta^2c_\theta^2 +  m_{\Phi}^2)\,.
\end{equation}
Notice, however, that upon the requirement  $m_\Theta>m_\Phi$, this leads to same constraint as in the ETC case,
Eq.~\eqref{eq:constraintCH}: $m_\Phi<m_\eta \ct$.
A more general discussion allowing a small mass for $\chi$ is provided in \app{sec:u1bosons}.

    \section{Dark Matter}
    \label{Sec:DM}
Before studying the phenomenology of the DM state, $\Phi$, it is illustrative to briefly discuss the relation between the TC and CH regimes.
In the TC limit the $\Pi_{UD}$ state in Table~\ref{table:symmetries} is stable due to the $\mathrm{U}(1)_{\rm TB} $ symmetry which is only violated by the EW anomaly above the EW scale.  
The kinetic term in the Lagrangian in Eq.~\eqref{eq:kinLag} includes contact interactions of $\Pi_{UD}$ with the SM gauge bosons of the form 
\begin{align}
\mathcal{L}_{\mathrm{kin}} \supset  -\frac{g^2}{2} s^2_{\theta} W_{\mu}^+W^{-\,\nu} \Pi_{UD}  \bar{\Pi}_{UD} 
\end{align}
which lead to a large thermal cross section
 \begin{align}
\langle  \sigma v\rangle = & \frac{g^4s^4_{\theta}}{128\pi  m_{\Pi_{UD}}^2}\sqrt{1-\frac{m_W^2}{m_{\Pi_{UD}}^2}}
\left(\frac{4m_{\Pi_{UD}}^4}{m_W^4}-\frac{4m_{\Pi_{UD}}^2}{m_W^2}+3\right) \nonumber\\
= & \frac{g^4s^4_{\theta} m_{\Pi_{UD}}^2}{ 32\pi m_W^4} +\mathcal{O}(m_W^2/m_{\Pi_{UD}}^2)\\
 \simeq  & 2 \cdot\, 10^{-24}  {\rm cm}^3/{\rm s}  \frac{s^4_{\theta} m_{\Pi_{UD}}^2}{m_W^2} \nonumber 
    \end{align}
In the TC limit where $s_{\theta}=1$, the $\Pi_{UD}$ state can therefore only be thermal dark matter if its mass is just below the $W$ mass threshold where the annihilation cross-section is kinematically suppressed~\cite{Belyaev:2010kp}. For $m_{\Pi_{UD}}>m_W$ the annihilation 
cross section is too efficient such that $\Pi_{UD}$ must instead be asymmetric if it is to be the DM as in~\cite{Ryttov:2008xe,Belyaev:2010kp}. In 
the CH parameter regime, with say  $s_{\theta}\lesssim 0.1$, the pNGB $\eta$, with a similar annihilation cross-section and with a weak scale mass, would be a WIMP candidate if it were stable but topological interactions make it unstable.

However the $\Phi$ state of the $\mathcal{R}_2$ sector in the extended models considered here remains protected 
by the $\mathrm{U}(1)_{\Lambda}$ symmetry in both the TC and CH vacua. The contact interactions of $\Phi$ with the SM gauge bosons 
is not only suppressed by $s_{\theta}$ in the CH vacuum, but also by the fact that the interactions arise only from the double trace terms in Eq.~\eqref{eq:intLag}.
    
In the following, we will compute the thermal annihilation cross sections for the three different cases for SM-fermion 
mass generation outlined in Sec.~\ref{sec:massMech}.

\subsection{Annihilation cross sections}

We list the dominant
annihilation cross sections for different channels below. For simplicity, we write down explicitly here only the channels
$\Phi\bar\Phi\to h_1h_1,VV$. Notice that the condition $m_\Theta>m_\Phi$ implies also $m_\eta>m_\Phi$ in all the cases 
of SM-fermion mass generation
that we consider; see Eqs~\eqref{eq:ThetaMassCH}, \eqref{eq:ThetaMassE}, and~\eqref{eq:ThetaMassPC}.
In the numerical analysis we keep all the interactions, and take into account 
the richer scalar sector in the pCH case including channels $\Phi\bar\Phi\to h_ih_j$, 
$i=1,2$.

\begin{equation}
    \label{eq:}
    \begin{split}
	\langle v\sigma\rangle_{h_1h_1}=&\frac{1}{16\pi s \fL^4}\sqrt{1-\frac{4m_{h_1}^2}{s}}
	    \left[s_\alpha^2 m_\Phi^2 g_{hh}+s_\alpha^2\frac{s-2m_\Phi^2}{2}d_{hh}+s_\alpha^2\frac{m_\Phi^2(s-2m_{h_1}^2)}{2}g_{\partial h}\right.\\
	    &\quad +s_\alpha^2\frac{s-2m_\Phi^2}{2}\frac{s-2m_{h_1}^2}{2}d_{\partial h}
	    -s_\alpha\left(m_\Phi^2 g_{h}+\frac{s-2m_\Phi^2}{2}d_h\right)\frac{g_{h_1h_1h_1}}{s-m_{h_1}^2}\\
	    &\left.\quad+c_\alpha\left(m_\Phi^2 g_{h}+\frac{s-2m_\Phi^2}{2}d_h\right)\frac{g_{h_2h_1h_1}}{s-m_{h_2}^2}
	    \right]^2 + \mathcal{O}(c_i^3),
    \end{split}
\end{equation}
and 
\begin{equation}
    \label{eq:}
    \begin{split}
    \langle v\sigma\rangle_{VV}=&\frac{\delta_V }{8\pi s\fL^4}\sqrt{1-\frac{4m_V^2}{s}}
	\left(3+\frac{s(s-4m_V^2)}{4m_V^4}\right)
	\left[m_\Phi^2 g_{VV}+\frac{s-2m_\Phi^2}{2}d_{VV}\right.\\
	&\quad-s_\alpha\left(m_\Phi^2 g_{h}+\frac{s-2m_\Phi^2}{2}d_h\right)\frac{g_{h_1VV}}{s-m_{h_1}^2}\\
	    &\left.\quad+c_\alpha\left(m_\Phi^2 g_{h}+\frac{s-2m_\Phi^2}{2}d_h\right)\frac{g_{h_2VV}}{s-m_{h_2}^2}	
	\right]^2,
    \end{split}
\end{equation}
for $V=W,Z$, and $\delta_{W,Z}=1,\frac{1}{2}$, resp. The couplings $g_i, d_i$, $i=h,\partial h, hh,VV$ 
are given explicitly in appendix~\ref{sec:intLag}, and the coupling $g_{h_{1}h_1h_1},g_{h_{1}VV}$, for CH and PC, and $g_{h_{i}h_1h_1},g_{h_{i}VV}$, $i=1,2$,
for pCH case can be obtained from the Lagrangians given above. The CH and PC cases correspond to $s_\alpha=-1$, see Eq.~\eqref{Eq:Higgsmixing}. 

Furthermore, the annihilation cross section $\Phi\bar\Phi\to t \bar{t}$ is given by
\begin{equation}
    \label{eq:}
    \begin{split}
    \langle v\sigma\rangle_{t\bar{t}}=\frac{3}{4\pi s \fL^4}\sqrt{1-\frac{4m_t^2}{s}}(s-4m_t^2)&
	\left[-s_\alpha\left(m_\Phi^2 g_{h}+\frac{s-2m_\Phi^2}{2}d_h\right)\frac{g_{h_1\bar{t} t}}{s-m_{h_1}^2}\right.\\
	&\left.\quad+c_\alpha\left(m_\Phi^2 g_{h}+\frac{s-2m_\Phi^2}{2}d_h\right)\frac{g_{h_2 \bar{t}t}}{s-m_{h_2}^2}
	    \right]^2,
    \end{split}
\end{equation}
where $g_{h_{1,2} \bar{t}t}$ are the couplings of the mass eigenstates $h_{1,2}$ to the top quark, and again
CH and PC cases correspond to $s_\alpha=-1$.

\bigskip
{\bf (i) ETC-type four-fermion operators.}
\begin{figure}
    \begin{center}
	\includegraphics[width=0.5\textwidth]{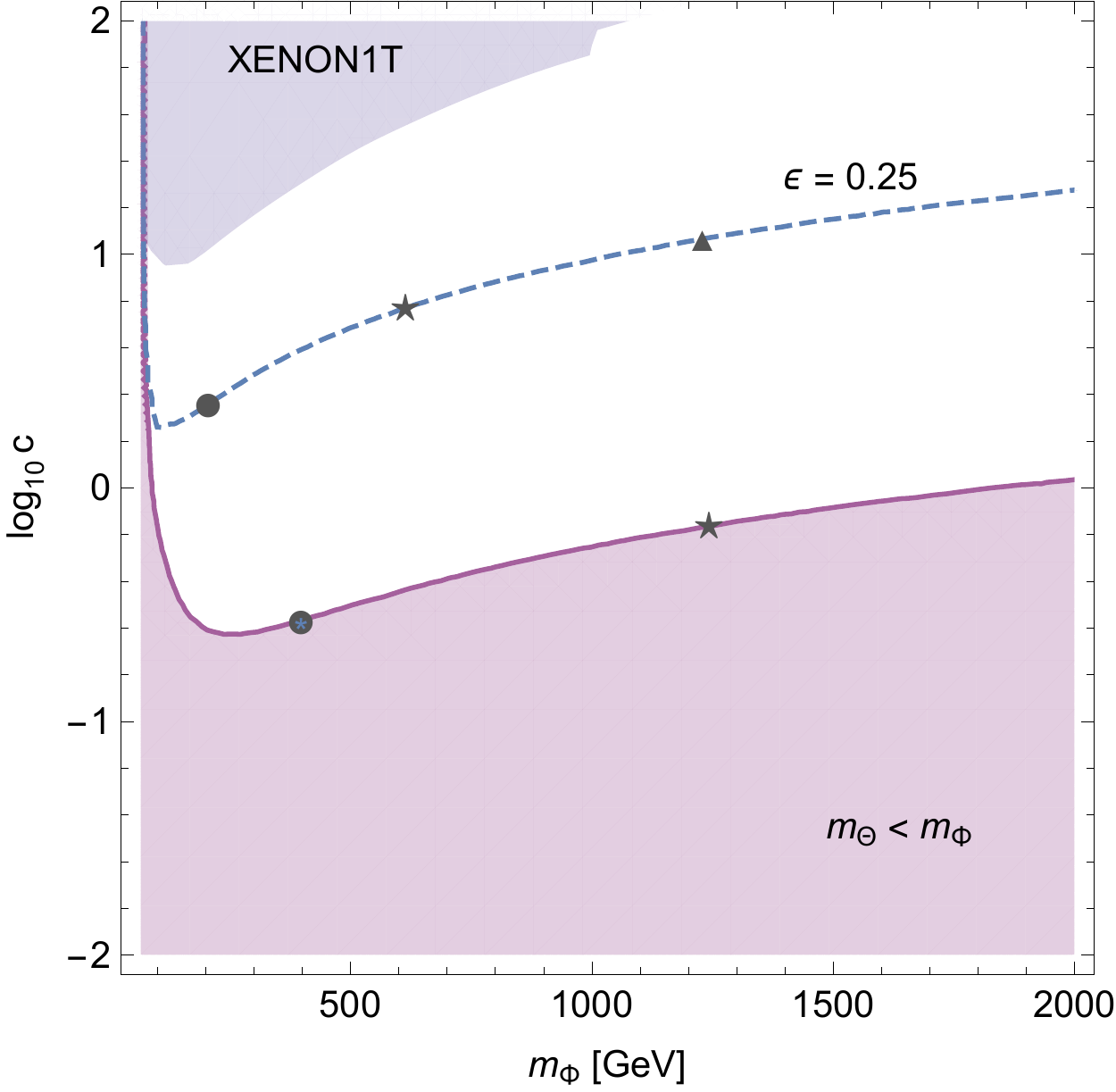}
    \end{center}
    \caption{The thermal cross section in the CH case for $c_1=c_2=c_3=c_4\equiv c$ 
	as a function of the DM mass, $m_{\Phi}$. We fix the DM mass as a function of compositeness scale, $f$, via
	$m_{\Phi}=\epsilon f$ and let $\epsilon$ vary. 
	The purple solid line corresponds to the thermal cross section 
	$\langle  \sigma v\rangle=3\cdot 10^{-26}\ \mathrm{cm}^3 \mathrm{s}^{-1}$ 
	for $\epsilon=\max\, \epsilon(m_\Phi)$ for which $m_\Theta=m_\Phi$. 
	We show $\max\, \epsilon(m_\Phi)$ in blue in Fig.~\ref{fig:epsMax}.
	In the shaded region below the purple solid line, the relic density is too small due to the annihilation 
	channel $\Phi \bar{\Phi} \to \Theta\Theta$ opening.  
	The upper shaded region shows the XENON1T exclusion~\cite{Aprile:2017iyp,Aprile:2018dbl} 
	for $\epsilon=\max\, \epsilon(m_\Phi)$; for smaller $\epsilon$, the region moves upwards, see 
	Eq.~\eqref{Eq:ddepsilon}.   The dot, star, and triangle correspond to $s_\theta=0.3,0.1,0.05$,
	respectively. Finally the blue dashed line shows the thermal cross section for a fixed $\epsilon=0.25$. 
	}
    \label{fig:thermalCH}
\end{figure}
We consider first the annihilation cross section in the purely CH case with ETC-type SM-fermion masses, described in Sec.~\ref{sec:massMech}. 
The condition $m_\Theta>m_\Phi$ 
(cf. Eq.~\eqref{eq:ThetaMassCH}) also implies $m_\eta>m_\Phi$, and therefore, the relevant annihilation channels are 
$\Phi\bar{\Phi}\to hh, VV, tt$.

Taking $m_{\Phi}=\epsilon f$, and $\fL=f=\vw/s_\theta$, and 
$m_Q=-\frac{m_h^2}{8\pi\cQ \vw\, t_\theta}$ fixed by the alignment condition Eq.~\eqref{eq:alignComp}, 
we find the leading-$\st$ contribution to the annihilation cross section to be 
\begin{equation}
    \label{eq:}
    \begin{split}
	\langle  \sigma v\rangle_{hh, VV}=\frac{4s^2_{\theta}\epsilon^2\left[32\pi\cQ \vw^2\epsilon^2 \left(c_2-16\pi\cL c_1 \right)
	    +m_h^2\left(c_3+16\pi\cL c_4\right)\right]^2}{(4\pi)^7 \cQ^2\cL^2\vw^6}+\mathcal{O}(s^4_{\theta},c_i^3).
    \end{split}
\end{equation}
The leading contribution from $\Phi\bar{\Phi}\to h \to  \bar{t}t$ is  $\mathcal{O}(s_{\theta}^4)$ and is given by
\begin{equation}
    \label{eq:}
    	\langle  \sigma v\rangle_{\bar{t}t}=\frac{3s^4_{\theta}m_h^4m_t^2\left(c_3+16\pi\cL c_4\right)^2}{(4\pi)^7 \cQ^2 \cL^2\vw^8}
	+\mathcal{O}(s_{\theta}^6).
\end{equation}

The relative contributions of $\langle  \sigma v\rangle_{\bar{t}t}$ and the two parts $\langle  \sigma v\rangle_{hh, VV}^{{1,2}}$ and $\langle  \sigma v\rangle_{hh, VV}^{{3,4}}$  from the $c_{1,2}$ and $c_{3,4}$ coefficients respectively 
are therefore
\begin{equation}
\label{Eq:relchannels}
\frac{\langle \sigma v \rangle_{\bar{t}t}}{\langle  \sigma v\rangle_{hh, VV}^{{3,4}}} =\frac{3 s_\theta^2}{4 \epsilon^2} \frac{m_t^2}{v_{w}^2}
 , \quad 
 \frac{\langle  \sigma v\rangle_{hh, VV}^{{3,4}}}{\langle  \sigma v\rangle_{hh, VV}^{{1,2}}} = \frac{1}{32 \pi c_Q \epsilon^2} \frac{m_h^2}{v_w^2} \frac{\left(c_3+16\pi\cL c_4\right)^2}{\left(c_2-16\pi\cL c_1 \right)^2}
\end{equation}
showing how the contact interactions, in particular the $VV$ interactions from the $c_{1,2}$ terms, dominate the annihilation cross-section parametrically in $\st$. These contact interactions are however loop-suppressed in direct-detection scattering.

Values of $\epsilon\sim\mathcal{O}(1)$ 
will quite naturally lead to
the right thermal relic density again with  $s_\theta\lesssim 0.1$, while 
the requirement $m_\Theta>m_{\Phi}$ sets an
upper bound $\epsilon\leq 1/2$. 

\begin{figure}
    \begin{center}
	\includegraphics[width=0.45\textwidth]{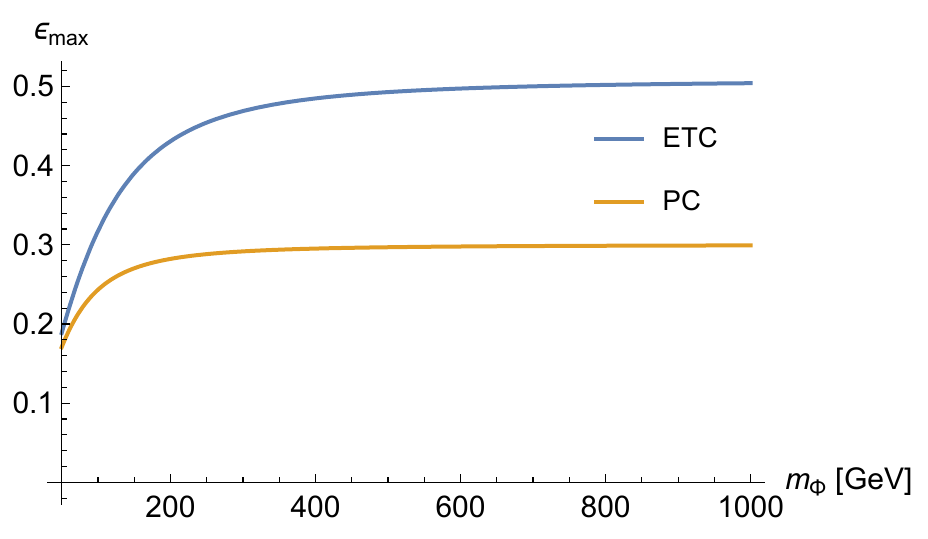}\quad
	\includegraphics[width=0.45\textwidth]{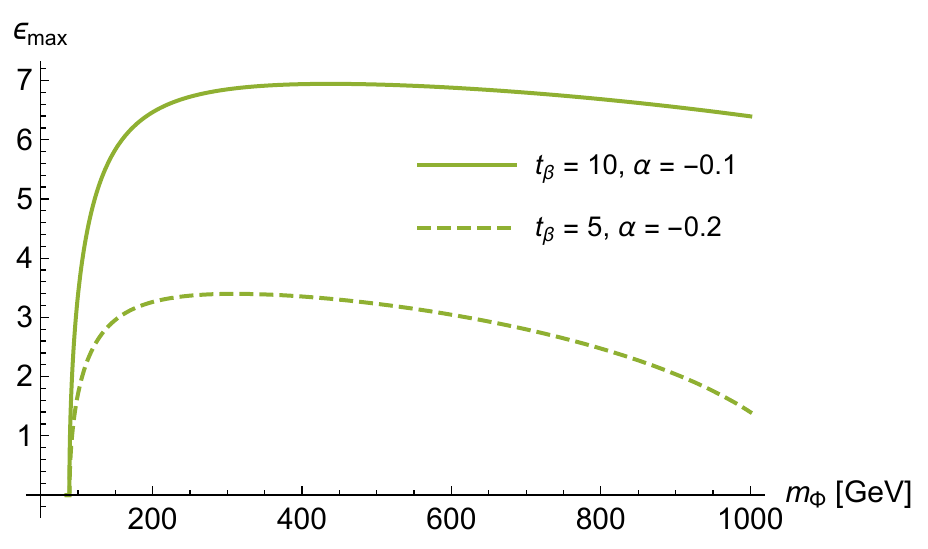}
    \end{center}
    \caption{The maximal value of $\epsilon=m_{\Phi}/f$ corresponding to $m_\Theta=m_\Phi$ in ETC and fermion PC scenarios (left panel), and for the benchmark cases ${t_\beta=10,\alpha=-0.1}$, ${t_\beta=5,\alpha=-0.2}$, in the partially composite Higgs scenario (right panel) as a function of $m_{\Phi}$.}
    \label{fig:epsMax}
\end{figure}

Setting $c_1=c_2=c_3=c_4\equiv c$, and $\cL=1$, we show the thermal cross section 
$\langle  \sigma v\rangle=3\cdot 10^{-26}\ \mathrm{cm}^3 \mathrm{s}^{-1}$ as a function of $c$ and $m_\Phi$ for different values 
of $\epsilon$ in Fig.~\ref{fig:thermalCH}. The blue dashed line corresponds to fixed value $\epsilon=0.25$, whereas on the purple solid 
line, $\epsilon=\max\, \epsilon(m_\Phi)$ corresponding to the limit $m_\Theta=m_\Phi$. 
We show $\epsilon=\max\, \epsilon(m_\Phi)$ in Fig.~\ref{fig:epsMax}.
On the shaded purple region $m_\Theta<m_\Phi$, and the relic abundance would be washed
away by the $\Phi\bar{\Phi}\to \Theta\Theta$ scatterings. The blue shaded region shows the current direct-detection limits
by XENON1T~\cite{Aprile:2017iyp,Aprile:2018dbl} assuming $\epsilon=\max\, \epsilon(m_\Phi)$. 
The direct-detection limits are discussed in more detail in Sec.~\ref{sec:DD}. 

\bigskip
{\bf (ii) Partially composite Higgs.}
Including an elementary Higgs doublet changes the picture significantly. Now CP-even states $\sigma_h$ and $h$ mix 
as described in Sec.~\ref{sec:massMech}, and the 
mass eigensates are the physical Higgs boson, $h_1$, and a heavy scalar, $h_2$.

In this case, the thermal annihilation cross section is determined by the scattering channels 
$\Phi\bar{\Phi}\to h_ih_j, VV, t\bar{t}$, where $i=1,2$. 
In the simplified case, 
where the additional scalar is heavy, and only $\Phi\bar{\Phi}\to h_1h_1,VV$ channels 
contribute, we can write the cross section as 
\begin{equation}
    \label{eq:txs-pCH}
    \begin{split}
    \langle  \sigma v \rangle =\frac{s_{\theta}^2\epsilon^6\left(16\pi\cL c_1+c_2\right)^2\left(s_{\beta}^4
	+3c_{\beta}^4\right)}{16\pi^5 \cL^2 c_{\beta}^2\vw^2}
	-\frac{s_{\theta}^4t_\beta^4\epsilon^4m_h^2\left(16\pi\cL c_1+c_2\right)\left(16\pi\cL c_4-c_3\right)}{256\pi^6 
	    \cL^2\cQ c_{\beta}^2\vw^4 \delta},
    \end{split}
\end{equation}
and we again take $m_{\Phi}=\epsilon f$, but 
now $\fL=f=\vw c_{\beta}/s_{\theta}$.  We fix $m_Q$ and $y_Q$ via the 
vacuum conditions, and trade $m_H^2$ and $\lambda_H$ for $\delta$ (defined in Eq.~\eqref{eq:alpha-delta}) and $m_h^2$.

\begin{figure}
    \begin{center}
	\includegraphics[width=0.45\textwidth]{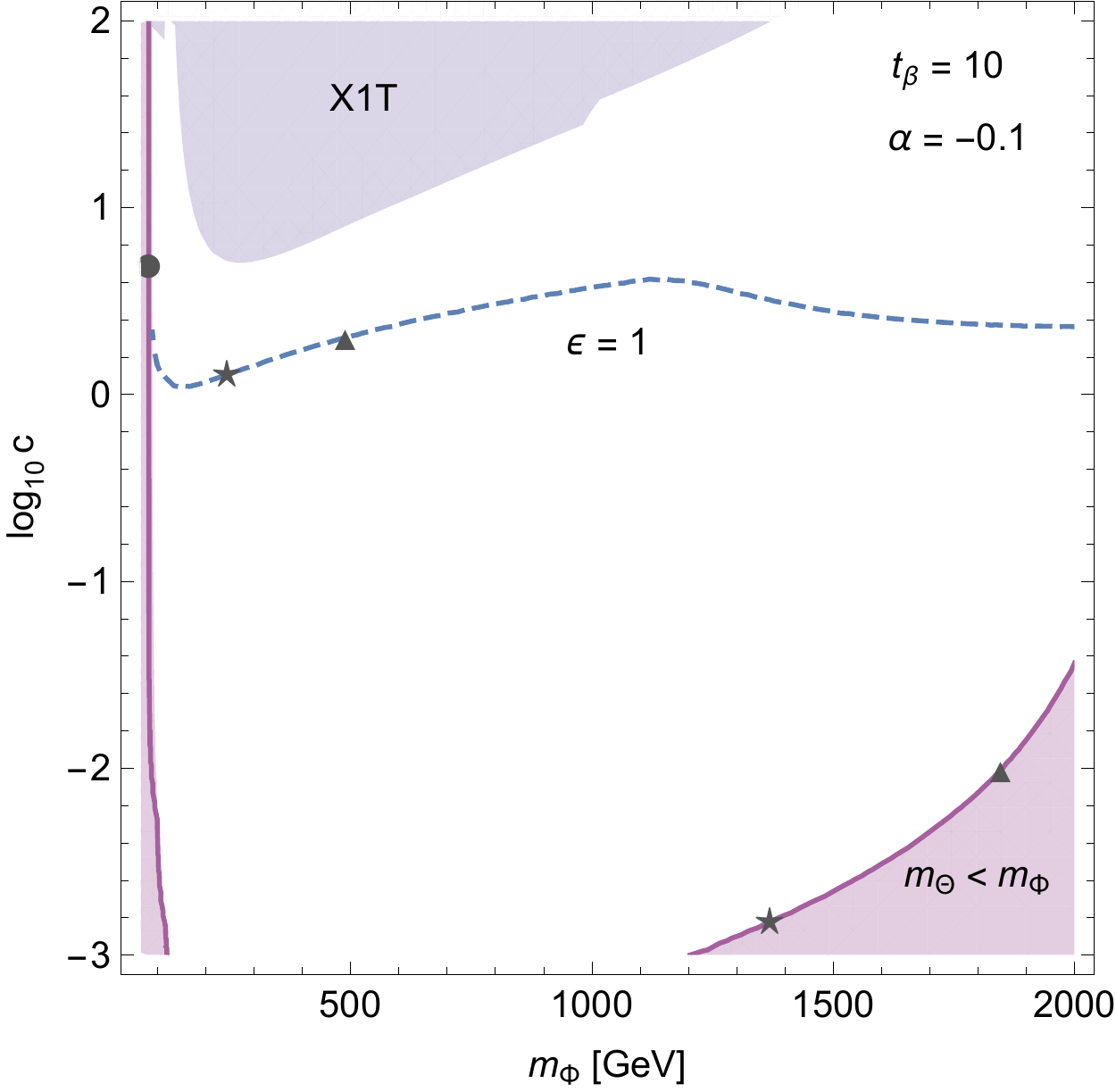}\quad
	\includegraphics[width=0.45\textwidth]{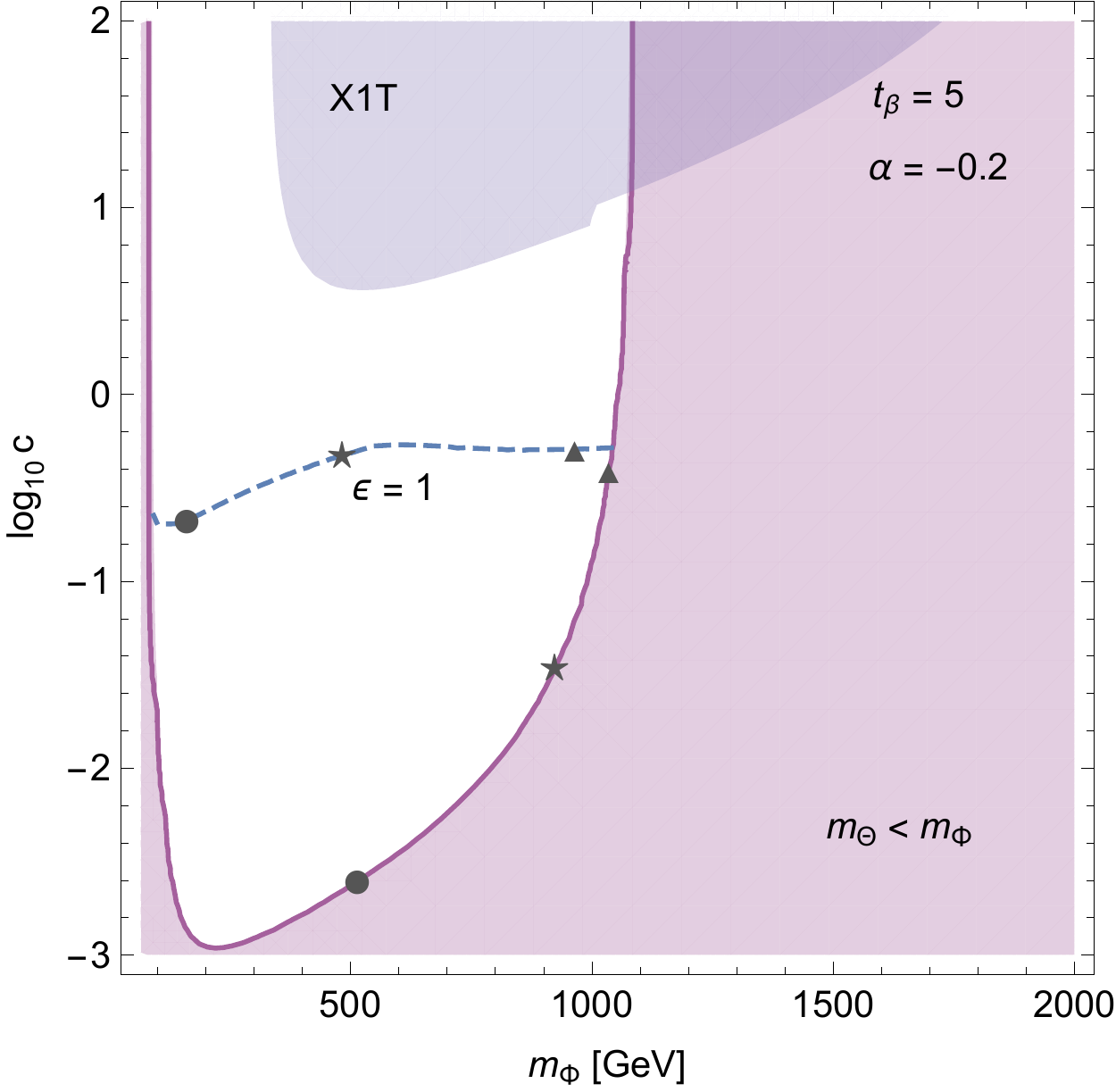}
    \end{center}
    \caption{The thermal cross section in the pCH case with $c_1=c_2=c_3=c_4\equiv c$ as a function of the 
    DM mass, $m_{\Phi}=\epsilon f$, for two benchmark cases: 
	$t_\beta=10, \alpha=-0.1$ (left panel) and  $t_\beta=5, \alpha=-0.2$ (right panel). 
	The blue dashed line corresponds the thermal cross section 
	$\langle  \sigma v\rangle=3\cdot 10^{-26}\ \mathrm{cm}^3 \mathrm{s}^{-1}$ 
	for fixed value $\epsilon=1$, whereas on the purple solid 
	line, $\epsilon=\max\, \epsilon(m_\Phi)$ corresponding to $m_\Theta=m_\Phi$; see Fig.~\ref{fig:epsMax}. 
	The upper shaded region shows the XENON1T exclusion~\cite{Aprile:2017iyp,Aprile:2018dbl} 
	assuming $\epsilon=\max\, \epsilon(m_\Phi)$; for smaller $\epsilon$, the region moves upwards.
	The dot, star, and triangle correspond to $s_\theta=0.3,0.1,0.05$,
	respectively.
    }
    \label{fig:thermalE}
\end{figure}

\begin{figure}
    \begin{center}
	\includegraphics[width=0.55\textwidth]{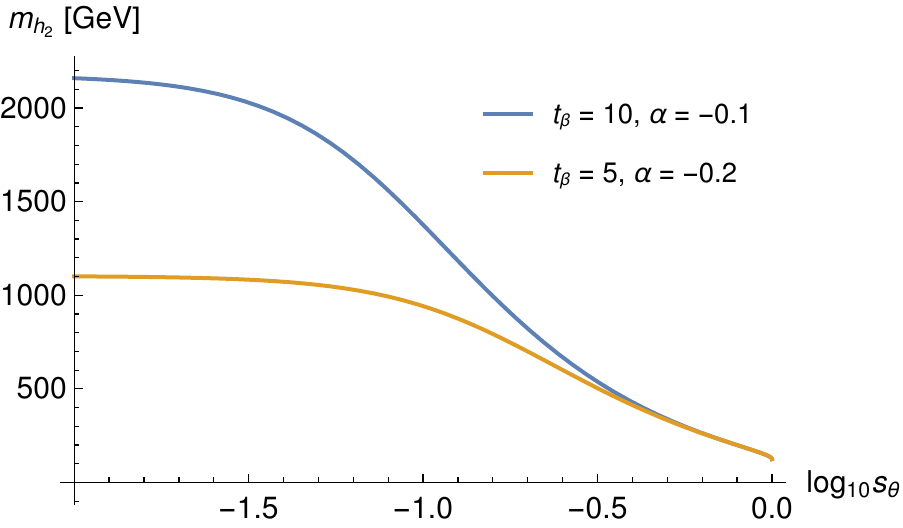}
    \end{center}
    \caption{The mass of $h_2$ as a function of $s_\theta$ for the cases shown in Fig.~\ref{fig:thermalE}. 
	The masses of $\eta$ and $\pi^{\pm,0}$ follow closely
	the mass of $h_2$ and we have omitted those for clarity.}
    \label{fig:spectE}
\end{figure}

We show the full result including the additional scalar channels, while again
setting ${c_1=c_2=c_3=c_4\equiv c}$ and $\cL=1$, in Fig.~\ref{fig:thermalE}
for two benchmark values of the addtional angles: (a) $t_\beta=10, \alpha=-0.1$ (left panel), and 
(b) $t_\beta=5, \alpha=-0.2$ (right panel). 
Again, the solid purple curve represents the thermal cross section
$\langle  \sigma v\rangle=3\cdot 10^{-26}\ \mathrm{cm}^3 \mathrm{s}^{-1}$ 
for $\epsilon=\max\, \epsilon(m_\Phi)$ 
corresponding to the limit $m_\Theta=m_\Phi$ (see Fig.~\ref{fig:epsMax}),
and the dashed blue curve corresponds to a fixed 
$\epsilon=1$. On the shaded purple region $m_\Theta<m_\Phi$,
 and upper shaded region shows the XENON1T exclusion~\cite{Aprile:2017iyp,Aprile:2018dbl} 
 for $\epsilon=\max\, \epsilon(m_\Phi)$. 
The associated heavy scalar spectrum corresponding to the benchmark cases (a) and (b) 
 is shown in Fig.~\ref{fig:spectE} as a function of $\st$. 
The kink in the thermal cross section lines are due to the
opening of the 
$\Phi\bar{\Phi}\to h_1h_2$ annihilation channel which yields the right relic abundance with lower values of $c$.

The top channel is now very subleading 
\begin{equation}
    \label{eq:}
    	\langle  \sigma v\rangle_{\bar{t}t}=\frac{3s^{8}_{\theta}s^8_\beta m_h^4m_t^2s_{\alpha-\beta}^4
	\left(c_3-16\pi\cL c_4\right)^2}{16\pi^7 s_{2\beta}^{10} \cQ^2 \cL^2\delta^2\vw^{8}}
	    +\mathcal{O}(s_{\theta}^{10}).
\end{equation}

\bigskip
{\bf (iii) Fermion partial compositeness.}
Taking $m_{\Phi}=\epsilon f$, and $f_{\lambda}=f=\vw/s_\theta$, and 
$m_Q$ fixed by the alignment condition Eq.~\eqref{eq:alignPC}, 
we find the corresponding cross section
including $\Phi\bar{\Phi}\to hh,VV$ channels ($\eta\eta$ channel is again excluded 
by the $m_\Theta>m_\Phi$ condition):
\begin{equation}
    \label{eq:}
    \begin{split}
	\langle  \sigma v\rangle=\frac{64s^2_{\theta}\epsilon^2}{(4\pi)^{11} \cL^2\cQ^2C_{yS}^4\vw^6}&\left[C_{tS}\left(16\pi\cL c_4 +c_3\right)
	    \left(8\pi^2m_t^2-C_{yS}^2\vw^2y_{t_R}^4\right)\right.\\
	    &\left.\quad-128\pi^3\cQ C_{yS}^2\vw^2\epsilon^2
	    \left(c2-16\pi\cL c_1\right)\right]^2
	    +\mathcal{O}(s^4_{\theta},c_i^3)
    \end{split}
\end{equation}
The leading contribution from $\Phi\bar{\Phi}\to \bar{t}t$ is  $\mathcal{O}(s_{\theta}^4)$ and is given by
\begin{equation}
    \label{eq:}
    	\langle  \sigma v\rangle_{\bar{t}t}=\frac{48C_{tS}^2m_t^2s^4_{\theta}
	\left(16\pi\cL c_4+c_3\right)^2\left(8\pi^2 m_t^2-C_{yS}^2\vw^2y_{t_R}^4\right)^2}{(4\pi)^{11} C_{yS}^4\cQ^2\cL^2\vw^8}
	+\mathcal{O}(s_{\theta}^{6}).
\end{equation}

\begin{figure}
    \begin{center}
	\includegraphics[width=0.5\textwidth]{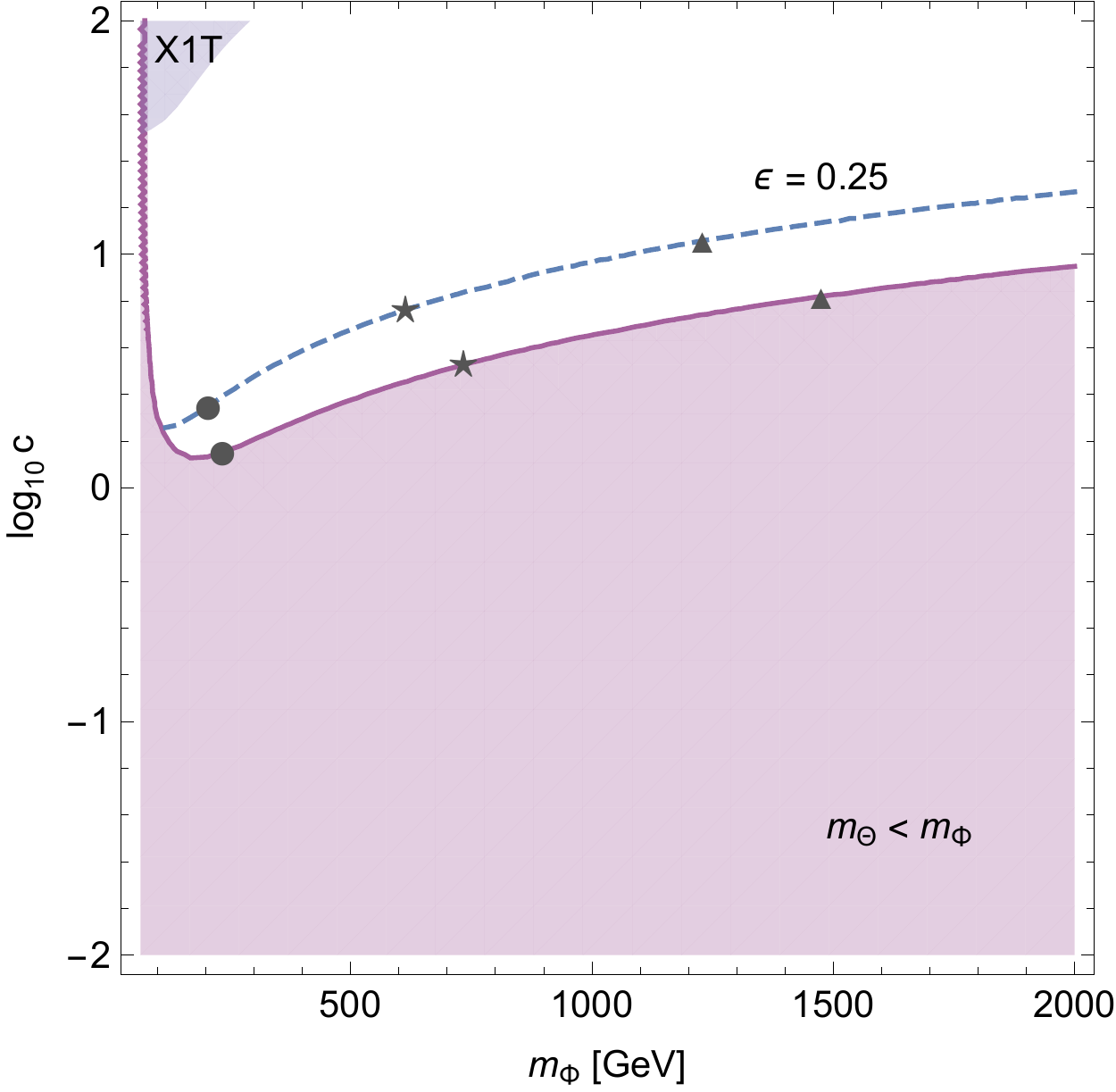}
    \end{center}
    \caption{The thermal cross section in the PC case for $c_1=c_2=c_3=c_4\equiv c$ as a function of 
    the DM mass, $m_{\Phi}=\epsilon f$. We have fixed $C_{yS}=10, C_{tS}=1$. 
    The blue dashed line corresponds the thermal cross section 
	$\langle  \sigma v\rangle=3\cdot 10^{-26}\ \mathrm{cm}^3 \mathrm{s}^{-1}$ 
	for fixed value $\epsilon=0.25$, whereas on the purple solid 
	line, $\epsilon=\max\, \epsilon(m_\Phi)$ corresponding to $m_\Theta=m_\Phi$; see Fig.~\ref{fig:epsMax}. 
	The upper shaded region shows the XENON1T exclusion~\cite{Aprile:2017iyp,Aprile:2018dbl} 
	assuming $\epsilon=\max\, \epsilon(m_\Phi)$.  The dot, star, and triangle correspond to $s_\theta=0.3,0.1,0.05$,
	respectively.
    }
    \label{fig:thermalPC}
\end{figure}

We show again the thermal cross section for $c_1=c_2=c_3=c_4\equiv c$, $\cL=1$, and $C_{yS}=10,C_{tS}=1$ 
as a function of $c$ and $m_\Phi$
in Fig.~\ref{fig:thermalPC}. The blue dashed line corresponds to fixed value $\epsilon=0.25$, and the purple solid 
line to $\epsilon=\max\, \epsilon(m_\Phi)$ for which $m_\Theta=m_\Phi$; see Fig.~\ref{fig:epsMax}. 
We note that the $\chi$ sector is assumed to be heavy. 

\subsection{Experimental searches}
\label{sec:DD}
The models we describe here 
can be experimentally searched for in multiple experiments including
(i) via underground direct-detection experiments searching for signals of DM scattering off nuclei, 
(ii) via direct production of the DM particles and other composite states or deviations in SM measurements at the LHC experiments, 
or (iii) indirectly via satellite missions looking for signals of 
DM annihilating into SM particles in the gamma-ray spectrum from astrophysical objects with high DM density.
We describe in this section the expectations for each of these types of searches.

\begin{figure}
    \begin{center}
    	\includegraphics[width=0.6\textwidth]{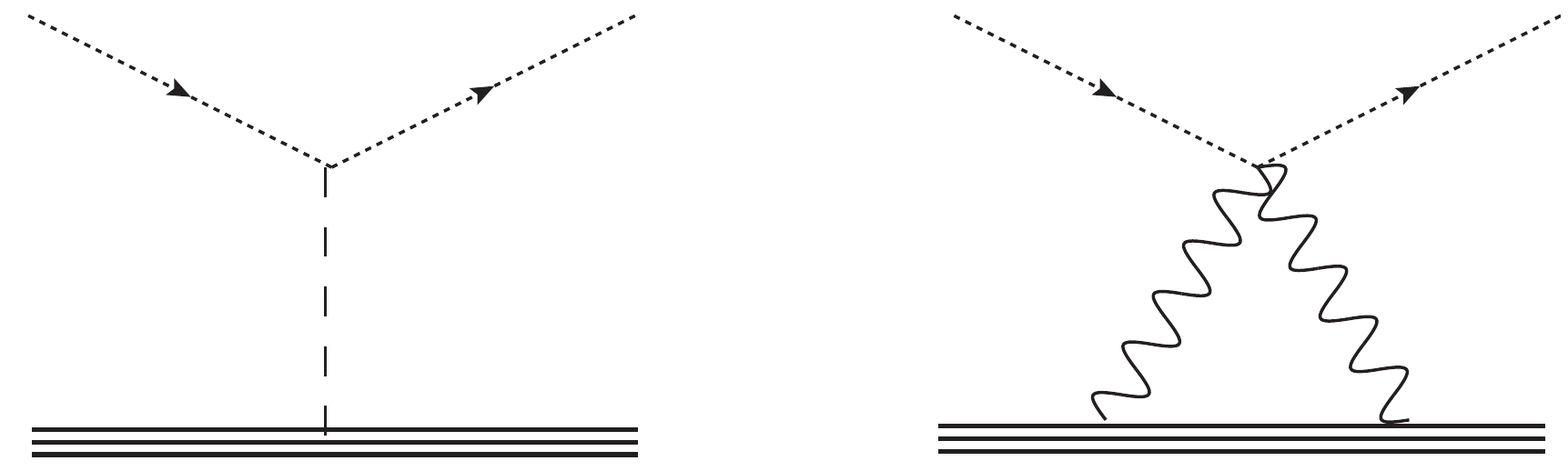} \\
    	(a) \hspace{4.5cm} (b)
    \end{center}
    \caption{\emph{Left:} Higgs exchange contribution to DD. \emph{Right:} WW loop.}
    \label{fig:diag}
\end{figure}

\vspace{0.5cm}
{\bf(i) Direct detection.} Since the contact interactions that determine the thermal relic density are loop suppressed in direct detection experiments, we expect the $t$-channel exchange of the (partially) composite scalars ($h_{1,2}$) $h$ (Fig.~\ref{fig:diag}(a)) will dominate the signal with a tree-level 
scattering cross-section on nuclei given by~\cite{McDonald:1993ex}:
\begin{equation}
\sigma_{\textrm{nucleon}}^H = \frac{\mu_N^2}{4\pi m_{\Phi}^2} 
\left[\frac{ g_{H}d_H  f_N m_N}{\vw}\right]^2, 
\end{equation}
where $\mu_N$ is the nucleon-DM reduced mass, $m_N$ the nucleon mass, $\vw$ the electroweak vev, $g_{H}$ is the 
effective coupling between the DM and the relevant (partially) composite Higgs, $H=h, h_1, h_2$,
$d_H$ describes the scalar exchange, 
and $f_N$ parametrizes the Higgs-nucleon coupling.

We take the central value of $f_N=0.3$~\cite{Alarcon:2011zs,Alarcon:2012nr} but this value depends on whether all fermion masses arise from the same mechanism or if e.g. only the top quark fermion mass does. 
The effective coupling $g_{H}$ is given in the limit of vanishing momentum transfer, $t\to 0$, by (cf. Eq.~\eqref{eq:intLag2}) 
\begin{equation}
\label{Eq:ddepsilon}
g_{H}= \frac{m_\Phi^2}{\fL^2}\left(g_{h} - g_{\partial h}\right) = \epsilon^2 \left(g_{h} - g_{\partial h}\right) \, ,
\end{equation}
where in the last equality we have used our assumption $f_\Lambda=f$. 
Depending on the SM-fermion mass mechanism, $d_H$ is given by 
\begin{equation}
    \label{eq:}
    \begin{split}
	d_H^{\mathrm{CH, PC}}&=\frac{1}{m_h^2},\\ 
	d_H^{\mathrm{pCH}}&= \frac{s_{\alpha}c_{\alpha}}{s_\beta}\left(- \frac{1}{m_{h_1}^2}+\frac{1}{m_{h_2}^2}\right).\\
    \end{split}
\end{equation}

The direct detection cross-section from the contact interactions are loop suppressed and the dominant ones from $W$ ($Z$) exhange are shown in Fig.~\ref{fig:diag}(b).  
Loop-induced direct-detection constraints from four-particle contact interactions between DM and SM photons and 
vectors bosons were evaluated in e.g. Refs~\cite{Frandsen:2012db,Crivellin:2014gpa} for interactions via 
field strength tensors and for inelastic transitions in Ref.~\cite{Cirelli:2005uq}.
Here however the interaction proceeds via the mass-like contact interaction, and we find the cross section
\begin{equation}
\sigma_{\rm nucleon}^{VV} = \frac{1}{16\pi}\left(\frac{(c_2-16\pi c_1 c_\Lambda)  g^2 g_{NV} m_N m_{\Phi} q}{64\sqrt{2} \pi ^2 \cL f^2 \fL^2}\right)^2 \, ,
\end{equation}
where $g_{NV}$ ($V=Z,W$) parametrizes the vector boson-nucleon couplings. For $\cL=1$, $q=1$ GeV, we get
\begin{equation}
\sigma_{\rm nucleon}^{VV}  \approx (8.0 g_{WN}^2+6.8 g_{ZN}^2)(c_2-16\pi c_1)^2 s_\theta^6\epsilon^2\,10^{-53} \text{ cm}^2.
\end{equation}

The resulting cross-section level is well below current direct-detection limits and below the neutrino floor. While in general there could be interference between the Higgs exchange and weak boson loop exchange, given how small the latter is, we do not consider that here.
In summary, the annihilation cross section is  dominated by the contact interactions of $\Phi$ with the 
SM vector bosons, while direct detection is dominated by the tree-level Higgs exchange.

\vspace{0.5cm}
{\bf (ii) The LHC experiments.} The DM  sector 
can be searched for at the LHC via missing energy signals. One class of such searches are the mono-$X$ channels with missing energy where $X$ is a jet, photon, a vector boson $V=Z,W$  or a Higgs boson~\cite{Birkedal:2004xn,Goodman:2010yf,Fox:2011pm}. 
The most distinct channels for our model are the $\bar{q}q$ initiated mono-$V$ channels, ${pp\to V^* \to V  \Phi \bar{\Phi}}$ 
through the quartic couplings in Eqs~\eqref{eq:intLag2b} and~\eqref{eq:intLag2bc}. The CMS and ATLAS experiments have looked at this kind of signature for a fermion DM candidate~\cite{Sirunyan:2017onm,Aad:2014vka} where dimension-5 operators are required.
These searches have been interpreted for scalar DM with a $g_{\mathrm{eff},V} V_\mu V^\mu\Phi \bar{\Phi}$ coupling in Ref.~\cite{Brivio:2015kia}  but only to derive an order-of-magnitude bound for the effective couplings. Our case corresponds to a good approximation to 
\begin{equation}
\label{eq:Zbound}
g_{\mathrm{eff},Z}= \frac{m_\Phi^2}{f_\lambda^2}(d_{ZZ}+g_{ZZ}) = \epsilon^2(g'^2+g^2)\st^2\left(\frac{c_2}{4\pi^2}+\frac{4c_1}{\pi}\right)\,,
\end{equation}
and currently only $g_{\mathrm{eff},Z}\gg 1$ is excluded~\cite{Brivio:2015kia}. For instance, in the ETC case where
the correct relic abundance can be obtained when $\epsilon\lesssim 0.5$ (see Fig.~\ref{fig:epsMax}),
$g_{\mathrm{eff},Z}= 1$ 
corresponds to $c\sim 80$ assuming $c=c_1=c_2$ and $\st=0.3$. 
Since the expression Eq.~\eqref{eq:Zbound}  has a similar dependence on $\theta$ and $\epsilon$ as the 
direct-detection bounds, even an optimistic bound on $g_{\mathrm{eff},Z}$ is at best comparable to the X1T limits. However, the inclusion of spin-one resonances can produce distinct $p_T$ spectra in these searches as shown in Ref.~\cite{Foadi:2008qv} and deserves to be investigated in detail.
 The monojet and monophoton limits on DM coupling to the SM via the Higgs portal, $pp\to H X \to \Phi \bar{\Phi} X$ are presented e.g. in Ref.~\cite{Balazs:2017ple}.  Limits on the coupling $g v h\Phi \Phi $ are $g\lesssim 1$ at $m_\Phi\simeq$ 100 GeV 
 and decrease at higher values of $m_\Phi$, but our equivalent coupling is orders of magnitude smaller.

The new strong sector can also be searched for directly at the LHC and future colliders. 
Depending on parameters the LHC might be able to directly produce the pNGBs $\eta$ and $\Theta$ of Table~\ref{table:symmetries} together with the additional $h_2$, $\pi^{\pm,0}$ in the pCH scenario (see Eq.~\eqref{Eq:Higgsmixing}) or the  $\Theta_2$ state in the PC scenario (see App.~\ref{sec:u1bosons}). 

A new aspect of our model, which has not yet been studied in detail,  is the appearance of the light $\Theta$ boson, which is a mixture of an inert and an EW charged component.
In addition, in the PC case we expect two further pNGBs as discussed above, which might lead to signatures not yet explored. Again they are here mixed with an inert sector 
and a careful analysis must be carried out. We leave this analysis for future work.
We note that in the PC case, other (probably heavier) QCD charged pNGB are present, and their phenomenology has been studied in Ref.~\cite{Cacciapaglia:2015eqa}.

The phenomenology of $\eta$ has been studied in Ref.~\cite{Arbey:2015exa} and the partially elementary scalars in Refs~\cite{Alanne:2017rrs,Alanne:2017ymh}. Furthermore, in Ref.~\cite{Alanne:2016rpe}, it was shown that the $\eta$ and $\eta'$  
corresponding to the quantum anomalous U(1) 
could be used to disentangle the three different fermion mass mechanisms we consider here.  
In Refs~\cite{Belyaev:2016ftv,Cacciapaglia:2017iws}, these bosonic states have been studied in a PC scenarios with both an EW sector 
and a QCD-colored sector 
but without inert fermions.

In summary, the model, and in particular its DM sector, leads to potentially rich collider phenomenology, 
but we leave the detailed study of it for future work.

\vspace{0.5cm}

{\bf (iii) Indirect detection.} Finally the Fermi-LAT limits from DM annihilations in dwarf spheroidals~\cite{Ackermann:2015zua} do not currently exclude a full thermal annihilation cross-section into $WW$ and $ZZ$ at any mass. This is in contrast to direct detection which is seen to do so, e.g.  in the right panel of Fig.~\ref{fig:thermalE}. We therefore do not include indirect detection in our limit plots. 
However, we note that 
with a order of magnitude improvement in the experimental sensitivity, the DM mass range $m_W \lesssim m_\Phi \lesssim $ 500 GeV can be probed.

\section{Conclusions}
\label{sec:conclusions}
In this paper we have constructed new (partially) composite Goldstone Higgs models featuring composite Goldstone DM candidates.  
The DM candidates are stabilized by $\mathrm{U}(1)$ global symmetries 
of the underlying four-dimensional gauge and gauge-Yukawa theories with strongly interacting fermions---analogous to the U(1) baryon symmetry responsible for the longevity of the proton. However, differently from the proton, the thermal relic density of the DM is particle anti-particle symmetric.

Only part of the strongly interacting fermions are gauged under the EW symmetry of the SM and the DM particle is the lightest particle charged under the global $\mathrm{U}(1)$ among the SM-inert fermions. At the effective Lagrangian level the DM relic density arises from double trace terms between the SM-inert-sector fermions and those that are gauged under the SM symmetry group. In particular these double trace terms provide four-point contact-interaction terms between the DM and the SM vector bosons---as well as the Higgs and additional pNGBs---which determine the thermal relic density. 

The dominant scattering channel for direct detection is via composite Higgs exchange arising also from double trace terms. The Higgs couplings to the DM particle is sensitive to the origin of the SM fermion masses, either via ETC-type four-fermion operators, mixing with elementary doublets, or via fermion partial compositeness. 
Therefore direct detection
is in principle able to probe this origin; however, the constraints are overall weak and only
exclude regions with unexpectedly large values of the Gasser--Leutwyler type coefficients
in the effective Lagrangian.

We eliminate the annihilation channel of the DM, $\Phi$, into the additional pNGB, $\Theta$, related to the global U(1) factor by requiring that $m_\Theta>m_\Phi$. This is the only annihilation channel that arises at single-trace level and which could wash out the thermal relic density.  This requirement selects vacuum angles in the region $\sin\theta \sim 0.01-1$ again depending on the fermion mass mechanism and assuming that the effective Lagrangian coefficients, $c_i$, are $\mathcal{O}(1)$ as expected by naive dimensional analysis. Therefore, the requirement eliminates the fine-tuned decoupling limit of parameter space where $\sin\theta \ll 1$.

The SM inert fermions may render the strong dynamics near-conformal and this may imply that the composite spin-one spectrum exhibits nearly parity doubling. In the TC limit this has been argued to reduce the electroweak $S$ parameter. Although in CH models with a large compositeness scale, $f$, such a dynamical reduction is not necessary to be in agreement with experimental constraints, it would be interesting to explore in partially composite Higgs models where the scale $f$ can be low. 
The presence of an explicit underlying model with composite Higgs and DM opens the possibillity for the lattice to provide crucial input to the phenomenology of these models such as a precise determination of the spectrum. 

Finally, the unique features of the model might present yet unexplored signatures at the LHC and future colliders.
     
\section*{Acknowledgments}
 MTF and MJ acknowledges partial funding from The Council For Independent Research, grant number 
DFF 6108-00623. The CP3-Origins centre is partially funded by the Danish National Research Foundation, grant number DNRF90.  DBF acknowledges partial funding by the European Union as a part of
the H2020 Marie Sk\l odowska-Curie Initial Training Network MCnetITN3 (722104).

\appendix

\section{$\Phi$-interactions } 
\label{sec:intLag}

Here we give explicitly the couplings in Eq.~\eqref{eq:intLag2},  
 \begin{equation}
 \label{eq:intLag2b}
 \begin{split}
\mathcal{L}_{Q,\Lambda}\supset &
\frac{m_{\Phi}^2}{\fL^2} \Phi\bar{\Phi} \left(g_h  h  + \frac{1}{2}g_{ZZ}   Z_\mu Z^\mu +  g_{WW}   W^+_\mu W^{-\mu} 
+  \frac{1}{2}g_{\Theta\Theta} \Theta^2  + \frac{1}{2} g_{hh}  h^2+ \frac{1}{2} g_{\eta \eta} \eta^2  \right.\\
    &\left.\qquad\quad+  \frac{1}{2}g_{\partial\Theta}\, \partial_\mu \Theta  \partial^\mu \Theta
	+  \frac{1}{2}g_{\partial h}\,  \partial_\mu h\partial^\mu h+  \frac{1}{2}g_{\partial\eta}\, \partial_\mu\eta\partial^\mu\eta \right)
  \\
  & +\frac{1}{\fL^2}\partial_\mu \Phi\partial^\mu \bar{\Phi} \left(   d_h  h  +\frac{1}{2} d_{ZZ}  Z_\mu Z^\mu +  d_{WW}   W^+_\mu W^{-\mu} 
    +  \frac{1}{2}d_{hh} h^2+ \frac{1}{2}d_{\eta \eta}  \eta^2 \right.\\
    &\left.\qquad\qquad\qquad+  \frac{1}{2}d_{\partial\Theta}\, \partial_\mu \Theta  \partial^\mu \Theta
	+  \frac{1}{2}d_{\partial h}\,  \partial_\mu h\partial^\mu h+ \frac{1}{2} d_{\partial\eta}\, \partial_\mu\eta\partial^\mu\eta \right).
  \end{split}
    \end{equation}
These read
\begin{equation}
 \label{eq:intLag2bc}
    \begin{aligned}[l]
	&g_h= \frac{c_3 m_Q \st}{2\cL\pi^2},\\
	&g_{ZZ}=-\frac{(g_{\mathrm{L}}^2+g_Y^2)c_2\st^2}{4\pi^2\cL},\\
	&g_{WW}=-\frac{c_2g_{\mathrm{L}}^2\st^2}{4\pi^2\cL},\\
	&g_{\Theta\Theta}=\frac{8\pi^2\cL\fL^2+4c_3m_Q f\ct}{9\pi^2\cL f_\Theta^2},\\
	&g_{\partial\Theta}= -\frac{8c_2}{9\pi^2\cL f_{\Theta}^2},\\
	&g_{hh}=\frac{c_3m_Q\ct}{2\pi^2\cL f},	\\
	&g_{\partial h}=-\frac{c_2}{\pi^2\cL f^2},\\
	&g_{\eta\eta}=\frac{c_3m_Q\ct}{2\pi^2\cL f},\\
	&g_{\partial \eta}= -\frac{c_2}{\pi^2\cL f^2},
    \end{aligned}
    \qquad\qquad
    \begin{aligned}[l]
	& d_h =\frac{8c_4m_Q\st}{\pi} ,\\
	& d_{ZZ} =\frac{4(g_{\mathrm{L}}^2+g_Y^2)c_1\st^2}{\pi} ,\\
	&d_{WW} =\frac{4g_{\mathrm{L}}^2c_1\st^2}{\pi} ,\\
	&d_{\Theta\Theta} = \frac{64c_4m_Q f\ct}{9\pi f_\Theta^2},\\
	&d_{\partial\Theta} = \frac{128c_1}{9\pi f_\Theta^2},\\
	& d_{hh} = \frac{8c_4m_Q\ct}{\pi f},\\
	& d_{\partial h} = \frac{16c_1}{\pi f^2},\\
	& d_{\eta\eta} = \frac{8c_4m_Q\ct}{\pi f} ,\\ 
	& d_{\partial\eta} =\frac{16c_1}{\pi f^2} 
    \end{aligned}
\end{equation}

\section{Abelian bosons}
\label{sec:u1bosons}

In a general underlying theory with $n$ fermionic sectors, each one specified by a representation $r$ and a number of fermions $N_r$, there is a global symmetry U(1)$^n$ one of which is anomalous under $G_{\mathrm{TC}}$ and the anomaly-free combinations are given by 
\begin{equation}
\sum_r q_r T(r)=0\,,
\label{eq:anomaly}
\end{equation} 
where $T(r)$ is the index of the representation $r$.
Each U(1) can be associated with a boson $\Theta_r$ and $n-1$ of them are pNGBs. 
The masses of these states can be parametrized by 
\begin{equation}
V_m = \sum_r\frac{1}{2}m_r^2 \Theta_r^2 + \frac{1}{2} m_A^2\Theta'^2\,,
\label{eq:u1masses}
\end{equation}
where $m_A$ is the mass generated by the anomaly and $\Theta'$ is the corresponding state. 
The states $\Theta_r$ can be parametrized inside the NGB matrices of \eq{eq:NGBmatrix} in the following way
\begin{equation}
    \Sigma_r = \exp\left[ 2\sqrt{2}\, i  \left(\frac{\Pi_r}{f_r}+\frac{\Theta_r}{\sqrt{2N_r}f_{\Theta_r}}\mathbbm{1}_{N_r}\right) \right] E_r, 
\end{equation} 
which defines canonically normalised kinetic terms from \eq{eq:kinLag}. $\mathbbm{1}_{N_r}$ is the identity matrix of dimension $N_r$. We proceed defining $n-1$ anomaly-free states
$\Theta_i$ ($i=1,\cdots, n-1$) and the anomalous $\Theta'\equiv \Theta_n$, with
\begin{eqnarray}
\Theta_r = \sqrt{2N_r}f_r\left( \sum_i q_{r,i} \frac{\Theta_i}{\mathcal{N}_i}   \right)\,,\quad\quad 
\mathcal{N}_i = \sum_r \sqrt{2N_r}f_r q_{r,i}\,.
\end{eqnarray} 
The factor $\mathcal{N}_i$ guarantees a proper normalization of the fields and 
\begin{equation}
\sum_i q_{r,i}q_{r',i} =0 
\end{equation}
guarantees no kinetic mixing.
 The charges of $\Theta_n$ that defines the anomalous combination are given by the vector perpendicular to the plane defined by the indexes, 
 \begin{equation}
 q_{r,n}=T(r)\,.
 \end{equation}
 The other charges are then defined to span the rest of the space with orthogonal basis.
These states  then mass mix via the terms in \eq{eq:u1masses} to form physical states. 

In ETC and pCH cases we assumed $n=2$ fermion representations, Q in {\bf F} and $\lambda$ in {\bf G} of $G_{\mathrm{TC}}=\mathrm{SU}(2)$, which give rise to 2 external U(1) groups one of which is anomalous. The anomalous charges are
$q_{F,2}=1/2$ and $q_{G,2}=2$ and the anomaly-free one is the orthogonal combination $q_{G,1}=1/2$ and $q_{F,1}=-1$. They mix according to \eq{eq:u1masses} and in the $\Theta'$ decoupled limit ($m_A\to \infty$) we recover \eq{eq:}: $m_\Theta^2=\frac{1}{9}(8m_F^2+m_G^2)$ with $m_F=m_\eta c_\theta$ and $m_G=m_\Phi$.  
A similar sitation with two sectors has been studied in Ref.~\cite{Belyaev:2016ftv}.

In PC case we assumed a $n=3$ fermionic sectors,  Q in {\bf F}, $\lambda$ in {\bf G} and $\chi$ in $\bf A_2$ of $G_{\mathrm{TC}}=\mathrm{Sp}(4)$. The anomalous combination is $q_{F,3}=1$, $q_{A_2,3}=3$, $q_{G,3}=3$. In the case the anomalous field is decoupled we have the masses of the two anomaly-free states given by
\begin{eqnarray}
m_{\Theta_1,\Theta_2}^2&=& \frac{1}{152}\left\{\phantom{\frac{}{}}23m_\Phi^2 +72m_F^2 + 52m_\chi^2 
  \pm \left[5184 m_F^4-432 m_F^2\left(5 m_{\Phi }^2+19 m_{\chi }^2\right) \right.\right. \nonumber  \\
  &+ & \left. \left.  529 m_{\Phi }^4+1710 m_{\Phi }^2 m_{\chi }^2+3249 m_{\chi }^4 \right]^{1/2} \right\}
\end{eqnarray}
with $m_F=m_\eta\ct$.
If the $\chi$ sector decouples ($m_\chi\to \infty$) then we recover the expression in \eq{eq:ThetaMassPC}:
$m_\Theta^2 =  \frac{1}{19}(18 m_\eta^2c_\theta^2 +  m_{\Phi}^2)$. If $m_\chi\ll m_F$ then $m_{\Theta_1}^2\approx\frac{1}{4}(m_\Phi^2+3m_\chi^2)$ and the phenomenology changes. This situation was not considered here and we leave it for future analysis.

%
\bibliography{PCH.bib}
\bibliographystyle{JHEP}
%
    
\end{document}